\documentclass[12pt,preprint]{aastex}
\usepackage{color}

\newcommand{\etal}{et\,al.}

\newcommand{\lsim}{\raise0.3ex\hbox{$<$}\kern-0.75em{\lower0.65ex\hbox{$\sim$}}}
\newcommand{\gsim}{\raise0.3ex\hbox{$>$}\kern-0.75em{\lower0.65ex\hbox{$\sim$}}}
\newcommand{\mjpbeam}{\,\,mJy\,beam$^{-1}$}
\newcommand{\msun}{M$_{\odot}$}
\newcommand{\hii}{H\,{\sc ii}}
\newcommand{\hi}{H\,{\sc i}}

\newcommand{\kms}{km\,s$^{-1}$}

\begin{document}

\slugcomment{\it Accepted for publication in the Astronomical Journal}

\title{ALFALFA Discovery of the Nearby Gas-rich Dwarf Galaxy Leo P. V.
  Neutral Gas Dynamics and Kinematics}


\author{Elijah Z. Bernstein-Cooper} 
\affil{Department of Astronomy, University of Wisconsin, 475 N
  Charter St Madison, WI 53706, USA;  Department of Physics \&
  Astronomy, Macalester College, 1600 Grand Avenue, Saint Paul, MN
  55105, USA} 
\email{ezbc@astro.wisc.edu}

\author{John M. Cannon}
\affil{Department of Physics \& Astronomy, Macalester College, 1600
  Grand Avenue, Saint Paul, MN 55105, USA}
\email{jcannon@macalester.edu}

\author{Edward C. Elson}
\affil{Astrophysics, Cosmology and Gravity Centre (ACGC),
  Department of Astronomy, University of Cape Town, Private Bag X3,
  Rondebosch 7701, South Africa}
\email{elson.e.c@gmail.com}

\author{Steven R. Warren}    
\affil{Department of Astronomy, University of Maryland, CSS Bldg.,
  Rm. 1024, Stadium Dr., College Park, MD 20742-2421, USA}
\email{swarren@astro.umd.edu}

\author{Jayaram Chengalur}
\affil{National Centre for Radio Astrophysics, TIFR, Post
  Bag 3, Ganeshkhind, Pune 411 007, India}
\email{chengalur@ncra.tifr.res.in}

\author{Evan D. Skillman} 
\affil{Minnesota Institute for Astrophysics, School of
  Physics and Astronomy, University of Minnesota, 116 Church St. SE,
  Minneapolis, MN 55455, USA}
\email{skillman@astro.umn.edu}
  
\author{Elizabeth A. K. Adams}
\affil{Center for Radiophysics and Space
  Research, Space Sciences Building, Cornell University, Ithaca, NY
  14853, USA}
\affil{Netherlands Institute for Radio Astronomy (ASTRON), Postbus 2,
  7990 AA, Dwingeloo, The Netherlands}
\email{betsey@astro.cornell.edu}

\author{Alberto D. Bolatto}    
\affil{Department of Astronomy, University of Maryland, CSS Bldg.,
  Rm. 1024, Stadium Dr., College Park, MD 20742-2421, USA}
\email{bolatto@astro.umd.edu}

\author{Riccardo Giovanelli}
\affil{Center for Radiophysics and Space
  Research, Space Sciences Building, Cornell University, Ithaca, NY
  14853, USA}
\email{riccardo@astro.cornell.edu}

\author{Martha P. Haynes}
\affil{Center for Radiophysics and Space
  Research, Space Sciences Building, Cornell University, Ithaca, NY
  14853, USA}
\email{haynes@astro.cornell.edu}

\author{Kristen B. W. McQuinn} 
\affil{Minnesota Institute for Astrophysics, School of
  Physics and Astronomy, University of Minnesota, 116 Church St. SE,
  Minneapolis, MN 55455, USA}
\email{kmcquinn@astro.umn.edu}

\author{Stephen A. Pardy} 
\affil{Department of Astronomy, University of Wisconsin, 475 N
  Charter St Madison, WI 53706, USA; Department of Physics \&
  Astronomy, Macalester College, 1600 Grand Avenue, Saint Paul, MN
  55105, USA} 
\email{spardy@astro.wisc.edu}

\author{Katherine L. Rhode}
\affil{Department of Astronomy, Indiana University, 727 East
  Third Street, Bloomington, IN 47405, USA}
\email{rhode@astro.indiana.edu,}

\author{John J. Salzer}  
\affil{Department of Astronomy, Indiana University, 727 East
  Third Street, Bloomington, IN 47405, USA}
\email{slaz@astro.indiana.edu}

\begin{abstract}

We present new \hi\ spectral line imaging of the extremely metal-poor,
star-forming dwarf irregular galaxy Leo\,P.  Our \hi\ images probe the
global neutral gas properties and the local conditions of the
interstellar medium (ISM).  The \hi\ morphology is slightly elongated
along the optical major-axis.  We do not find obvious signatures of
interaction or infalling gas at large spatial scales.  The neutral gas
disk shows obvious rotation, although the velocity dispersion is
comparable to the rotation velocity.  The rotation amplitude is
estimated to be V$_{\rm c}$ $=$15\,$\pm$\,5 km\,s$^{-1}$.  Within the
\hi\ radius probed by these observations, the mass ratio of gas to
stars is roughly 2:1, while the ratio of the total mass to the
baryonic mass is $\gsim$15:1.  We use this information to place Leo\,P
on the baryonic Tully-Fisher relation, testing the baryonic content of
cosmic structures in a sparsely populated portion of parameter space
that has hitherto been occupied primarily by dwarf spheroidal
galaxies.  We detect the signature of two temperature components in
the neutral ISM of Leo\,P; the cold and warm components have
characteristic velocity widths of 4.2\,$\pm$\,0.9 \kms\ and
10.1\,$\pm$\,1.2 km s$^{-1}$, corresponding to kinetic temperatures of
$\sim$1100 K and $\sim$6200 K, respectively.  The cold \hi\ component
is unresolved at a physical resolution of 200 pc.  The highest
\hi\ surface densities are observed in close physical proximity to the
single \hii\ region.  A comparison of the neutral gas properties of
Leo\,P with other extremely metal-deficient (XMD) galaxies reveals
that Leo\,P has the lowest neutral gas mass of any known XMD, and that
the dynamical mass of Leo\,P is more than two orders of magnitude
smaller than any known XMD with comparable metallicity.

\end{abstract}						

\keywords{galaxies: evolution --- galaxies: dwarf --- galaxies:
  irregular --- galaxies: individual (Leo\,P)}

\section{Introduction}
\label{section: intro}

The exploration of the neutral interstellar medium (ISM) of
star-forming, low-mass galaxies has become a major observational and
theoretical focus of extragalactic astrophysics.  Motivated by the
long-standing ``missing satellites'' problem
\citep{kauffmann93,klypin99,moore99} and by recent extensions of the
baryonic Tully-Fisher relationship to low rotation velocities
\citep[e.g.,][]{stark09,mcgaugh10a,mcgaugh12}, major interferometric
survey programs have been undertaken to characterize the morphology
and dynamics of individual gas-rich dwarfs in the Local Volume
(THINGS, {Walter \etal\ 2008}\nocite{walter08}; FIGGS, {Begum
  \etal\ 2008b}\nocite{begum08b}; SHIELD, {Cannon
  \etal\ 2011}\nocite{cannon11a}; VLA-ANGST, {Ott
  \etal\ 2012}\nocite{ott12}; LITTLE THINGS, {Hunter
  \etal\ 2012}\nocite{hunter12}).  Based on the results of these
surveys, star-forming galaxy disks with neutral hydrogen masses
exceeding $\sim$10$^{7}$ \msun\ are numerous, well-studied, and
comparatively well-understood.

In contrast, the parameter space of the most extremely low-mass and
gas-rich systems remains only sparsely investigated.  The review of
\citet{mcconnachie12} lists 12 systems in or near the Local Group with
detected neutral hydrogen masses M$_{\rm HI}$ $<$10$^{6}$ \msun\ (dSph
or dE/dSph systems including Sculptor, Fornax, NGC\,205, and NGC\,185;
``transition'' objects classified as dIrr/dSph including LGS\,3,
Phoenix, Leo\,T, UGC\,4789, ESO\,410-G\,005, and ESO\,294-G\,010; and
dIrr systems including Antlia and KKH\,86).  Together, the
aforementioned major surveys add a few more systems with M$_{\rm HI}$
$<$10$^{6}$ \msun\ (LGS\,3 was observed in the {LITTLE THINGS
  survey}\nocite{hunter12}, and Antlia, KDG\,73, KKH\,86, and KK\,230
were observed in the {VLA-ANGST project}\nocite{ott12}).  Similarly,
selected individual systems outside of the Local Group have also been
discovered recently: \citet{roychowdhury12} catalog dJ0926$+$7030 and
dJ1012$+$64 (which appears to be UGC\,5497 via cross-listing in the
NASA Extragalactic Database) as blue compact dwarf systems in the
M\,81 group.  To our knowledge, this is the complete sample of
presently known galaxies with detected neutral hydrogen masses
$<$10$^{6}$ \msun\ (excluding LeoP, the subject of this paper).  To
date, none of these systems has been the subject of a detailed mass
distribution analysis.

One of the primary goals of the Arecibo Legacy Fast ALFA (ALFALFA) blind
extragalactic \hi\ survey is to populate the faint end of the \hi\ mass
function with statistical confidence \citep{giovanelli05}.  Major follow-up
survey programs targeting ALFALFA detections are underway, sampling low-mass,
\hi-rich systems that both possess and lack obvious stellar counterparts in
moderate-depth optical survey data products: SHIELD (Cannon \etal\
2011)\nocite{cannon11a} and the ``ultra-compact high velocity cloud''
(``UCHVC'') sample (Adams \etal\ 2013)\nocite{adams13}.  Given the extremely
faint optical luminosities of most of the objects in this mass range, the
latter initiative may be an especially promising avenue by which to discover
new low-mass galaxies in the Local Volume.  The challenge is that accurate
distances must also be established through detection of their associated
stellar populations.  As discussed in \citet{adams13}, if optical imaging
reveals a stellar component and a distance within $\sim$1 Mpc, then dozens of
the candidate UCHVCs could in fact be nearby dwarf galaxies.

Recently, the ALFALFA survey discovered the galaxy known as Leo\,P
from its \hi\ signature alone \citep{giovanelli13}.  The
\hi\ structural parameters of Leo\,P are consistent with it being
classified as a UCHVC according to the criteria in \citet{adams13}.
Subsequent optical imaging \citep{rhode13} and spectroscopy
\citep{skillman13} have shown that this system is among the most
extreme star-forming systems known in the local universe (see
Table~\ref{table: properties} for a summary of physical parameters).
Located at a distance of 1.72$^{\rm +0.14}_{\rm -0.40}$ Mpc (i.e.,
outside but in the immediate vicinity of the Local Group; {McQuinn
  \etal\ 2013}\nocite{mcquinn13}), this system harbors both young and
old stellar populations, as well as a single \hii\ region whose
nebular oxygen abundance (measured by the direct line method in {Skillman
  \etal\ 2013}\nocite{skillman13}) is only 3\% Z$_{\odot}$; this is
equal within the uncertainties to that of I\,Zw\,18 (Skillman \&
Kennicutt 1993)\nocite{skillman93} and DDO\,68 {(Berg
  \etal\ 2012)}\nocite{berg12}.  The combination of the very small
total luminous baryonic mass (M$_{\rm HI}$ $=$ 9.1\,$\times$\,10$^{5}$
\msun; M$_{\star}$ $=$ 5.7\,$\times$\,10$^{5}$ \msun), the presence of
ongoing star formation, the extremely low oxygen abundance, and the
relative isolation (the nearest known neighbor, Sextans\,B, is
0.47$^{+0.14}_{-0.24}$ Mpc away in 3D space; see details in {McQuinn
  \etal\ 2013}\nocite{mcquinn13}), make Leo\,P unique among nearby
dwarf galaxies. It represents an ideal opportunity to understand
numerous facets of galaxy evolution, including the nature of star
formation in the metal-poor ISM, the phase structure of the ISM, the
dynamics of extremely low-mass gaseous disks, and the baryonic content
of low-mass halos.

In this fifth paper in the series of manuscripts about Leo\,P (papers
one through four are {Giovanelli \etal\ 2013}\nocite{giovanelli13},
{Rhode \etal\ 2013}\nocite{rhode13}, {Skillman
  \etal\ 2103}\nocite{skillman13}, and {McQuinn
  \etal\ 2013}\nocite{mcquinn13}, respectively; paper six by {Warren
  \etal, in preparation, presents CARMA CO imaging of Leo\,P), we
  study the neutral gas morphology and dynamics of Leo\,P.  We
  organize this manuscript as follows.  In \S~\ref{section:
    observations} we describe the data acquisition, reduction and
  analysis.  \S~\ref{section: gas+stars+dark} presents the
  \hi\ morphology and kinematics, while \S~\ref{section: rotation}
  discusses the rotation of Leo\,P.  In \S~\ref{section: cold hi} we
  derive the properties of the cold and warm ISM in Leo\,P.
  \S~\ref{section: context} places Leo\,P in context amongst other
  metal-poor galaxies.  In \S~\ref{section: conclusion} we present our
  conclusions.

\section{Observations and Data Handling}
\label{section: observations}

We acquired three sets of \hi\ 21-cm spectral-line observations with the
\textit{Karl G. Jansky Very Large Array} \footnote{The National Radio
  Astronomy Observatory is a facility of the National Science
  Foundation operated under cooperative agreement by Associated
  Universities, Inc.} (VLA). C-configuration (VLA/C) and
B-configuration (VLA/B) observations were acquired in April and August
of 2012, respectively, for Director's Discretionary Time program
VLA/12A-456 (AC1105; PI Cannon). D-configuration (VLA/D) observations
were acquired in March of 2013 for program VLA/13A-026 (AC1114; PI
Cannon). We centered an IF-band on 1419.04 MHz and used a bandwidth of
1,000 kHz (4,000 kHz for VLA/D) with 256 channels (1,000 for VLA/D),
resulting in a spectral channel separation of 3.9 kHz (0.83
km\,s$^{-1}$\,ch$^{-1}$).  The primary calibrator was 3C286 and the
phase calibrator was J12021+2159. Total on-source integration times
amounted to 5 hours in the B-configuration, and 3 hours each in the C-
and D-configurations.  Unfortunately, the B-configuration observations
were compromised by a class M5.5 solar flare that occurred on August
18, 2012\footnote{http://www.swpc.noaa.gov}; Leo\,P was less than
9\arcdeg\ from the Sun during these observations.

\textit{Giant Metrewave Radio Telescope} (GMRT) observations of Leo\,P
were conducted in December of 2012, amounting to 20.5 hours of
on-source integration time. We implemented a similar observational
set-up to the VLA observations, the primary difference being that the
GMRT observations had a spectral resolution of 8.13 kHz (1.72
km\,s$^{-1}$\,ch$^{-1}$).

We calibrated each dataset individually via standard prescriptions using
CASA\footnote{Common Astronomy Software Application (CASA) is developed and
maintained by the NRAO.} and AIPS\footnote{Astronomical Image Processing
System (AIPS) is developed and maintained by the NRAO.} for the VLA
observations. The initial calibration and flagging of the GMRT data was done
using the flagcal package \citep{prasad12,chengalur13}. We subtracted the
continuum from each $uv$-dataset using a first-order linear fit leaving only
the line source.  We smoothed all datasets to a common spectral channel
separation of 2.5 \kms\,ch$^{-1}$ by using CASA to shift all input databases
to the same spectral grid in the Local Standard of Rest frame.  We then
concatenated all of the data into a single $uv$ database; all $uv$ weights
were set to unity during concatenation.

Next, we inverted and imaged all four $uv$-datasets together in the
AIPS environment.  We used a Gaussian weighting scheme of shorter and
longer baselines in the $uv$-plane to produce four cubes with beam
areas increasing by a factor of four.  We also produced a cube with
24\arcsec\ (200 pc) resolution to facilitate direct comparison with
previous work (see detailed discussion in \S~\ref{section: cold
  hi}). We performed residual flux rescaling on each of the cubes in
order to account for the different shapes of the clean beam and the
dirty beam \citep{jorsater95}. These cubes were spatially convolved
with elliptical Gaussian functions to produce circular beam sizes of
4\arcsec, 8\arcsec, 16\arcsec, 24\arcsec, and 32\arcsec; these cubes
have root-mean-square (RMS) noises (measured in line-free channels of
the non-rescaled cubes) of 0.45, 0.40, 0.51, 0.64, and 0.66 \mjpbeam,
respectively.

We derived moment maps and the global profile for each of the cubes by
following the THINGS protocols described in \citet{walter08}. We began
by spatially convolving the cube to twice the resolution and then
blanking at the 2.5$\sigma$ level in the new cube. Next, we
hand-blanked this cube and used it as a mask to blank the original
cube. We integrated all emission in the masked original resolution
cube to create the moment zero map. Moment one and moment two maps
were derived by including all emission in the masked original
resolution cube above the 5$\sigma$ level.  The integrated
\hi\ spectrum was derived from the masked original resolution cube.
All velocity information presented is calculated in the Local Standard
of Rest frame. \hi\ column density maps were calculated assuming
optically thin \hi\ gas.\footnote{For optically thin \hi\ gas, N$_{\rm
    HI}$ = 1.823 $\times 10^{18} \int T_b d\nu$ cm$^{-2}$ where $T_b$
  is brightness temperature and $\nu$ is frequency.} Note that second
moment maps made from datacubes blanked in this way provide lower
limits to the velocity dispersions, particularly in regions of low
signal-to-noise.

\section{Gaseous, Stellar, and Dark Components}
\label{section: gas+stars+dark}
\subsection{Neutral Gas Morphology}
\label{section: neutral gas morphology}

In Figure~\ref{fig: chmaps} we present individual channel maps of the
\hi\ emission at 16\arcsec\ resolution.  \hi\ emission is detected
across roughly 40 km\,s$^{-1}$ of velocity space ($\sim$240 \kms\ -
280 \kms).  The full (non-blanked) cube is shown in order to convey
the relative signal to noise ratio in each channel of the cube.  There
is a signature of rotation in these panels, evidenced by the position
of the \hi\ maximum in each panel moving slightly from northwest to
southeast at progressively higher recessional velocities.

We create global \hi\ profiles for Leo\,P by summing the flux in each
of the blanked datacubes and plotting versus velocity in
Figure~\ref{fig: spectra}.  The left panel compares the
32\arcsec\ resolution profile with those derived from the ALFALFA data
products as well from the follow-up confirmation observation with the
L-band Wide receiver at Arecibo (from which an \hi\ global flux
integral of 1.31\,$\pm$\,0.04 Jy\,km\,s$^{-1}$ is derived; see
{Giovanelli \etal\ 2013}\nocite{giovanelli13} for details).  There is
good agreement of these profiles, both in terms of the line profile
shapes and in terms of the recovered flux integrals.  The right panel
shows the global profiles at each angular resolution.  The global flux
densities are 1.03, 0.93, 0.65 and 0.31 Jy\,km\,s$^{-1}$ from the
32\arcsec, 16\arcsec, 8\arcsec, and 4\arcsec\ cubes, respectively.  As
expected, due to masking, the profiles become narrower and recover
less flux at successively higher angular resolutions.

The systemic velocity (V$_{\rm sys}$) of Leo\,P is estimated using the
methods of \citet{springob05}, as applied in \citet{pardy14}. We
identify the peak flux, $f_p$, and then fit a first-order polynomial
to the rising side of the spectrum using all fluxes between 15\% and
85\% of $f_p$. We define V$_r$ as the velocity where the fitted
polynomial equals 50\% of $f_p$. We repeat the fit for the falling
side of the spectrum, and define V$_l$ in the same way as V$_r$. The
velocity width, W$_{\rm 50}$, is defined as the difference between the
velocities V$_l$ and V$_r$. The systemic velocity, V$_{\rm sys}$, is
then the average of V$_l$ and V$_r$ weighted by the flux.  Leo\,P has
a systemic velocity of V$_{\rm sys}$ $=$ 260.8 $\pm$ 2.5 \kms\ in the
Local Standard of Rest Kinematic (LSRK) frame and a full-width
half-maximum of W$_{\rm 50}$ $=$ 22.5 $\pm$ 3.7 \kms.

\hi\ moment zero images, representing \hi\ column density or \hi\ mass
surface density, are presented in Figures~\ref{fig: coldens},
\ref{fig: lbt}, and \ref{fig: lbt zoom}.  In Figure~\ref{fig:
  coldens}, the \hi\ moment zero images at each angular resolution are
shown in greyscale, with contours overlaid.  These same contours are
then overlaid on the optical broad-band image from the Large Binocular
Telescope (LBT\footnote{The LBT is an international collaboration
  among institutions in the United States, Italy, and Germany. LBT
  Corporation partners are: The University of Arizona on behalf of the
  Arizona university system; Istituto Nazionale di Astrofisica, Italy;
  LBT Beteiligungsgesellschaft, Germany, representing the Max-Planck
  Society, the Astrophys- ical Institute Potsdam, and Heidelberg
  University; The Ohio State University, and The Research Corporation,
  on behalf of The University of Minnesota, The University of Notre
  Dame, and The University of Virginia.}) in Figure~\ref{fig: lbt};
this optical image was originally presented in \citet{mcquinn13}.
Similarly, in Figure~\ref{fig: lbt zoom}, the \hi\ moment zero image
at 4\arcsec\ resolution is compared with the same LBT image as in
Figure~\ref{fig: lbt}, as well as with the H$\alpha$ image from the
KPNO 2.1m telescope from \citet{rhode13}.

At low resolution, the peaks in the HI column density correspond to
peaks in the stellar surface brightness.  Interestingly, as one moves
outward from the optical and \hi\ center of the system, there are
stars that are associated with Leo P (see the individual star lists in
both {Rhode \etal\ 2013}\nocite{rhode13} and in {McQuinn
  \etal\ 2013}\nocite{mcquinn13}) that are beyond the regions in which
we detect \hi\ gas (most noticeably in the northwest region of the
galaxy).  Most of these outlying stars to the northwest are older red
giant stars; the blue main-sequence stars in Leo\,P are concentrated
in the region occupied by the high surface density gas in
Figure~\ref{fig: lbt}.  There is also a low surface brightness
extension of the \hi\ gas to the southeast, inside of which few stars
associated with Leo\,P are detected.  We do not interpret these as
signatures of tidal interaction, since we see no concrete evidence of
\hi\ gas on larger spatial scales when the $uv$ data are imaged with a
Gaussian taper to produce spatial resolutions larger than
32\arcsec\ (further, recall the discussion of the relative isolation
of Leo\,P in \S~\ref{section: intro}).  A deep single-dish
\hi\ mapping mosaic of the region surrounding Leo\,P would provide
more stringent constraints on the nature of any tidal interaction
Leo\,P may have undergone; ALFALFA mapping has not revealed any
gas-bearing feature in the vicinity of Leo\,P.

A considerable amount of the \hi\ gas in Leo\,P lies in central
regions of high surface-brightness \hi.  Moving to progressively
higher spatial resolutions, the \hi\ becomes increasingly localized
around the highest surface brightness stellar components of the
galaxy.  As Figure~\ref{fig: lbt zoom} shows, the highest \hi\ column
densities (N$_{\rm HI}$ $\gsim$ 4\,$\times$\,10$^{20}$ cm$^{-2}$,
corresponding to an \hi\ mass surface density $\sigma_{\rm HI}$ $>$
3.2 \msun\,pc$^{-2}$) are mostly co-spatial with the youngest massive
stars identified in \citet{rhode13} and in \citet{mcquinn13}.  Two
regions reach a peak \hi\ column density of 6.5\,$\times$\,10$^{20}$
(5.2 \msun\,pc$^{-2}$); one is nearly cospatial with the single
\hii\ region, while a second peak is located slightly offset from the
main stellar component.  This latter peak is not currently associated
with ongoing star formation.

\citet{molter14} and \citet{warren14} report sensitive new
\textit{Combined Array for Research in Millimeter-wave Astronomy}
(CARMA) CO (1-0) observations of Leo\,P.  A detailed comparison of
those CO upper limits with the present \hi\ and optical data will
allow unique tests of models of star formation in the extremely
metal-poor ISM.  We defer such analyses to \citet{warren14}.

\subsection{Neutral Gas Kinematics}
\label{section: neutral gas kinematics}

The 16\arcsec\ resolution \hi\ moment zero images are compared with
the moment one (intensity-weighted velocity field) and with the moment
two (\hi\ velocity dispersion) images in panels (a), (c), and (e) of
Figure~\ref{fig: pvslice maps}.  The moment one map reveals an obvious
but low-amplitude ($\lsim$15 km\,s$^{-1}$) velocity gradient from the
northwest to the southeast regions of Leo\,P.  However, the moment one
map is relatively noisy and does not show a simple velocity field;
significant random motions are superposed on the bulk rotation
signature.  This is unlike the regular rotation seen in some more
massive dIrr galaxies \citep[e.g.,][]{ott12,hunter12}.

Part of the disorder of the velocity field shown in Figure~\ref{fig:
  pvslice maps} can be accounted for by Leo\,P's velocity
dispersion. The moment two map (also shown in Figure~\ref{fig: pvslice
  maps}) indicates that the velocity dispersion is $\gsim$8
\kms\ throughout the disk of Leo\,P.  Superposed on this narrow
dispersion gas is a component with velocity dispersions of 10-12 \kms.
These velocity dispersion values are comparable to those seen in other
dwarf galaxies (see, e.g., the discussion in {Warren
  \etal\ 2012}\nocite{warren12} and {Stilp
  \etal\ 2013}\nocite{stilp13}, and references therein).  Leo\,P is
similar to some of the lowest-mass gas-rich galaxies studied in
\citet{begum03}, \citet{begum04}, \citet{begum08b}, \citet{ott12}, and
references therein, where the average velocity dispersion across the
disk is comparable to the estimates of the magnitude of the coherent
rotational velocity (see \S~\ref{section: rotation}).

In order to get a robust estimate of the global velocity dispersion,
we fitted Gaussian functions to individual profiles in the
datacube. We first spatially binned the 16\arcsec\ resolution cube to
6\arcsec\ pixels and then fit spectral profiles including emission
above 5$\sigma$ with both one Gaussian and two Gaussians. Double
Gaussian fits converged to only a handful of pixels; these results are
discussed in \S~\ref{section: cold hi} as evidence for two thermally
stable phases of the atomic gas. We thus estimate the global velocity
dispersion using only single Gaussian fits; the average of the
resulting Gaussian fits is 8.4 \kms, and the standard deviation of the
Gaussian widths is 1.4 \kms. We thus conclude that the global velocity
dispersion of Leo\,P is 8.4 $\pm$ 1.4 \kms.

A first-order characterization of the rotation of Leo\,P is available
by extracting spatially-resolved position-velocity (PV) slices through
the 16\arcsec\ resolution datacube.  This strategy is similar to the
one presented in \citet{cannon11b}.  The kinematic major-axis is
identified as that axis along which a PV slice gives the largest
projected velocity gradient. This angle, measured counterclockwise
from north to the approaching side of the disk, is 335\arcdeg; this
angle is obvious from the velocity fields presented in
Figure~\ref{fig: pvslice maps} and is in agreement with the apparent
optical major-axis (see Figures~\ref{fig: lbt} and \ref{fig: lbt
  zoom}, and the images presented in {Rhode
  \etal\ 2013}\nocite{rhode13} and {McQuinn
  \etal\ 2013}\nocite{mcquinn13}).  We extract PV slices through the
datacubes, summing all emission in beam-wide areas, each separated by
a full beam width.  Slices are extracted along both the kinematic
major-axis and the kinematic minor-axis.  This produces multiple PV slices
through the datacube, each of which is independent from its
neighbor. The central slice along each axis is centered on the
\hi\ column density maximum.  The orientations of these slices are
shown by arrows overlaid on the moment zero, one, and two images at
16\arcsec\ resolution in Figure~\ref{fig: pvslice maps} (b), (d), and
(f), respectively.

The spatially-resolved PV slices are shown in Figure~\ref{fig:
  pvslices} for the 16\arcsec\ cube.  The number of each panel
corresponds to the number of the slice in Figure~\ref{fig: pvslice
  maps}; the major-axis slices are shown in the upper panel, and the
  minor-axis slices are shown in the lower panel. The direction of the
  slices is such that a positive angular offset (in Figure~\ref{fig:
    pvslices}) corresponds to movement in the direction of the arrows
  shown in Figure~\ref{fig: pvslice maps}.

From these PV slices we draw the following conclusions.  First, the
velocity extent of the galaxy is comparable along the major-axis and
the minor-axis, and is only slightly larger than the central velocity
dispersion of the gas (i.e., the projected rotation velocity is of
order the velocity dispersion).  Second, the minor-axis cuts allow a
coarse estimate of velocity and position at different ends of the
disk; if this is interpreted as rotation, it sets a lower bound for
the rotation velocity (since the PV slices apply no inclination
correction).  The observed difference in these velocities is $\lsim$20
km\,s$^{-1}$. Taking half of this as the projected rotation, we
estimate the rotation amplitude to be $\sim$10 km\,s$^{-1}$ or less;
we discuss this amplitude further in \S~\ref{section: rotation}.
Third, there is evidence for localized concentrations of \hi\ with
modest velocity widths on the order of the velocity dispersion.  These
are apparent in both the major-axis and the minor-axis slices, most
obviously in the central cuts through the \hi\ column density maxima.
Fourth, the extension of \hi\ to the southeast appears as an asymmetry
in the PV slices (compare cuts \#1 and \#5 on the minor-axis; this is
also evident in major-axis cut \#2).

\section{The Rotation of Leo\,P}
\label{section: rotation}

Rotation curve work using \hi\ spectral line observations requires
both high spectral and high spatial resolution data (i.e., many beams
across the disk, with velocity resolution much finer than the
magnitude of the velocity gradient across neighboring beams), as well
as a galactic disk that has a favorable geometric orientation.
Knowing various parameters such as kinematic major-axis angle,
systemic velocity, dynamical center position, and inclination, one can
fit a series of tilted rings to the observed velocity field that is
representative of a thin rotating galactic disk.  Such
well-constrained rotation curves allow for a detailed analysis of the
baryonic-to-dark matter ratio as a function of radius, in addition to
constraining the total dynamical mass within the outermost ring.

We were unable to fit a single unambiguous rotation curve to the
velocity field of Leo\,P at any angular resolution without enforcing
priors.  While the kinematic major-axis angle, systemic velocity, and
velocity dispersion are well-constrained by the data, a range of
values for the dynamical center, inclination, and rotation velocity
were able to return adequate fits to the data.  In order to extract
dynamical information about Leo\,P from these data, we thus adopt the
value of inclination, $i$ $=$ 60\arcdeg, from \citet{giovanelli13}.
This inclination is based on both the \hi\ and the optical axial
ratios.  Below we use this value of inclination to estimate the
dynamical mass of Leo\,P.

We have two methods with which to estimate the rotation velocity of
Leo\,P.  The first is to exploit the spatially-resolved PV diagrams
shown in Figure~\ref{fig: pvslices}.  As discussed in \S~\ref{section:
  neutral gas kinematics}, the minor-axis PV slices allow one to
measure the projected velocity at either end of the disk.  Assuming
that the difference in these velocities is caused by the rotation of
the system, this provides an estimate of the rotation amplitude,
V$_{\rm rot}$ $\lsim$ 10 km\,s$^{-1}$.  Assuming $i$ $=$
60\arcdeg\ (see above), the inclination correction of $\simeq$15\%
increases the rotation velocity by less than $\sim$2 \kms.

The second and more robust method with which to estimate the rotation
velocity of Leo\,P is a major-axis PV slice.  The positions of the
spatially-resolved PV slices (width equal to the beam size) are shown
in Figure~\ref{fig: pvslice maps}, and the resulting slices are shown
in Figure~\ref{fig: pvslices}.  These panels reveal a total velocity
extent of 10$\pm$5 \kms\ for gas above the 3$\sigma$ level, within
$\pm$60\arcsec\ of the assumed dynamical center.  It is important to
note that these major-axis PV slices only include pixels summed over a
single beam width.

From these major-axis PV slices, we determine a maximum velocity
offset of $\pm$10 \kms\ from systemic, over an angular separation of
$\pm$60\arcsec.  Specifically, the maximum projected velocity of
$\sim$270 \kms\ occurs in the extreme southeast region of the disk,
while the minimum projected velocity of $\sim$250 \kms\ occurs in the
northwest region of the disk.  Along the direction of the PV slice,
the maximum offset between these locations is roughly $-$1\arcmin\ to
$+$1\arcmin\ (corresponding to $\pm$500 pc).  This is slightly smaller
than the linear extent over which \hi\ is detected; a line with
position angle $=$ 335\arcdeg, passing through the \hi\ column density
maximum of the 16\arcsec\ moment zero map (see Figure~\ref{fig:
  pvslice maps}) and traversing the entire detected \hi\ extent, has a
length of 184\arcsec.

Having constrained the total rotation velocity, we now consider
effects that may bear on the interpretation of these results.  The
first of these is the possible contribution of pressure support to the
disk.  Given the comparable magnitudes of the projected rotation
(V$_{\rm rot}$ $\sim$ 10 \kms; see above) and the velocity dispersion
(8.4 $\pm$ 1.4 \kms\ throughout the disk, with significant regions of
the system showing values of 10-12 \kms; see Figure~\ref{fig: pvslice
  maps} and discussion in \S~\ref{section: neutral gas kinematics}),
it is likely that random motions in the neutral gas provide some level
of dynamical support within the disk of Leo\,P.

Traditionally, the effects of pressure support within a galactic disk
are modeled by the asymmetric drift correction.  One parameterization
of this correction \citep{frusciante12} is given by

\begin{equation} 
    {\rm V^2_{c}(r)=
        V^2_{rot}(r) - \sigma^2(r)\ \left( \frac{d\,log \ \mu(r)}{d\,log \ r} +
        \frac{d\,log \  \sigma^2(r)}{d\,log\  r} \right)}
\label{eq: asym drift} 
\end{equation}

\noindent where V${\rm _c}$ is the true rotation velocity or circular
velocity (corresponding to the true dynamical mass in the absence of
pressure support) in \kms, V$_{\rm rot}$ is the observed gas rotation
velocity in \kms, $\sigma$ is the velocity dispersion of the gas in
\kms, $\mu$ is the gas mass surface density in \msun\,pc$^{-2}$, and r
is the radius in pc.  The values of $\mu$ and $\sigma$ are usually
measured by averaging emission in the moment zero and moment two maps
(using the tilted ring parameters derived from rotation curve
analysis), fitting the resulting data points with a differentiable
function, and evaluating at a given radius. Since we do not have an
unambiguous rotation curve for Leo\,P, we simply used the GIPSY task
ELLINT to create radially-averaged profiles of the \hi\ surface
brightness and of the \hi\ velocity dispersion.  The magnitude of the
resulting asymmetric drift correction reaches $\sim$5 \kms\ at a
radius of 64\arcsec\ (530 pc).  As expected based on the comparable
magnitudes of the projected rotation velocity and the velocity
dispersion, the dynamical support provided by random motions is
appreciable.

The second possible effect is beam smearing. The observed rotation
amplitude can underestimate the true amplitude if the beam size
smoothes over a significant gradient in velocity.  To test for the
possibility of beam smearing in our data, we used the GIPSY task
GALMOD to construct model datacubes with maximum rotation amplitudes
of 10 \kms\ and 15 \kms; the inclination was fixed at 60\arcdeg.  We
then smoothed these cubes to 16\arcsec\ and 32\arcsec\ resolutions.
In each case, we are able to differentiate between a 10 \kms\ rotator
and a 15 \kms\ rotator that has been severely beam smeared.  We thus
conclude that the effects of beam smearing are minor in these data.

Based on the above discussion, we conclude that the circular velocity
of Leo\,P is 15\,$\pm$\,5 \kms.  While evidence favors a velocity less
than 10 \kms, the inclination of the source and the significant
contribution of pressure support to the disk may allow the true
rotation amplitude to reach 15 \kms.

Perhaps the most critical physical parameters that can be derived from
the datasets in hand are the total dynamical mass (within the
outermost radius within which \hi\ is detected) and the luminous
baryonic mass.  Using the angular distance of 60\arcsec\ (r $=$ 500 pc
at 1.72 Mpc) discussed above for the maximum measured radial distance
within the disk, and conservatively assuming a circular velocity of
V$_{\rm c}$ $=$ 15 \kms\ at this radius, a simple \begin{math}
  \frac{V_{c}^2{\cdot}r}{G}\end{math} calculation gives a total
dynamical mass estimate of M$_{\rm dyn}$ $>$2.6\,$\times$\,10$^{7}$
\msun. This value is quoted as a lower limit because it only includes
the mass within the \hi\ radius; it seems likely that the dark matter
halo extends well beyond this radius.  The luminous baryonic mass is
considered to be the sum of the gas mass and the stellar mass.  We
implicitly assume that the dust and H$_{\rm 2}$ contents of Leo\,P are
modest compared to these quantities; this is in agreement with
measurements of the dust mass of the extremely metal poor galaxy
I\,Zw\,18 \citep{cannon02,herreracamus12,fisher14}.  We also assume
that the ionized gas mass in Leo\,P is negligible compared to the mass
of neutral hydrogen.  The total gaseous mass (M$_{\rm HI}$ $+$ 35\%
for Helium) of Leo\,P is found to be 1.2\,$\times$\,10$^{6}$
\msun\ (applying the Arecibo value of S$_{\rm HI}$ $=$ 1.31 Jy\,\kms).
As noted above, M$_{\rm \star}$ $=$ 5.7\,$\times$\,10$^5$ \msun.  This
yields a total baryonic mass of M$_{\rm bary}$ $=$ (1.8$^{\rm
  +0.4}_{\rm -1.2}$)\,$\times$\,10$^6$ \msun, where the asymmetric
error bars reflect those on the distance measurement from
\citet{mcquinn13}.

\section{The Cold \hi\ in Leo\,P}
\label{section: cold hi}

In the first paper in our Leo P series, \citet{giovanelli13}
postulated that a thermally stable multi-phase neutral ISM may be
present in Leo P. They suggested that these phases, if present, have
temperatures of T $\lsim$ 1,000 K for the cold phase and T $\gsim$
6,000 K for the warm phase. A multi-phase ISM has important
implications for models of the ISM in dwarf galaxies whereby the
relative masses of the two components may help govern the amount of
H$_{\rm 2}$ formation and the available pressure support
\citep[e.g.,][]{pelupessy06,faerman13}.

To test the hypothesis that two thermally stable \hi\ phases are
present in Leo\,P, we adopted the methods presented in \citet{warren12}
to search for a cold neutral medium. The method seeks to separate the
contribution of the two components in the ISM to the overall spectrum
at any given location in a galaxy.  For full details of the
identification process we refer the reader to \citet{warren12};
briefly, we first produced image cubes with linear resolutions of 130,
200, and 265 pc (16\arcsec, 24\arcsec, and 32\arcsec; see below for
our justification of these scales). Next we extracted the spectrum
from each 1.5\arcsec\ pixel and fit both a single and double Gaussian
function to those spectra with peak-to-rms S/N $>$ 10. The residuals
of each fit were then compared via a single-tailed F-test and a double
Gaussian was deemed a significantly better fit to the data if the
probability of improvement was at the $>$95\% confidence level.

Our choice of linear scales in which to search for the cold \hi\ is
driven by both data and physical reasons. \citet{warren12} showed that
in order to reliably separate multiple Gaussian components one needs
to have data with S/N $>$ 10. Our 4\arcsec\ and 8\arcsec\ cubes have
few pixels that fulfill this basic criterion. Also, in order to
separate the cold \hi\ component from the warm \hi\ component, there must
be a strong enough cold \hi\ signal. Thus, lower resolution actually
helps this procedure. However, if one uses too large a convolution,
there arises a risk of diluting the cold \hi\ signal if the beam's linear
scale is much larger than the scales in which the cold \hi\ exists by
way of including more of the ubiquitous warm \hi\ component in the
spectrum.

Where the cold \hi\ component traces molecular gas, comparisons to
current star formation tracers in the literature show that
correlations between $\Sigma$(H$_{\rm 2}$) and $\Sigma$(SFR) break
down on scales below $\sim$100 pc (e.g., {Onodera
  \etal\ 2010}\nocite{onodera10}). If the molecular clouds in Leo\,P
have similar separation distances as those in the Milky Way ($\sim$200
pc; {Koda \etal\ 2006}\nocite{koda06}), then we would expect to
include the flux density of multiple clouds on linear scales of at
least this size. For all of the above reasons we fit 16\arcsec,
24\arcsec, and 32\arcsec\ cubes (130, 200, and 265 pc linear scales,
respectively). The 24\arcsec\ $=$ 200 pc cube has the added benefit of
allowing us to directly compare to the results in \citet{warren12}, as
they worked at this spatial scale.

We detect the signature of cold \hi\ in the same area in Leo\,P at
each of these spatial scales. The 16\arcsec\ $=$ 130 pc detections are
slightly resolved as the cold \hi\ is found at 3 independent
pixels. The 24\arcsec\ $=$ 200 pc and 32\arcsec\ $=$ 265 pc cold
\hi\ detections are unresolved. Figure~\ref{fig: cold hi spectrum}
shows the spectrum at the location of the peak cold \hi\ component
detected in the 200 pc linear resolution data. The cold \hi\ component
represents 30\% of the total \hi\ flux density at this peak location,
which is comparable to the typical 20\% flux density contribution seen
in \citet{warren12}.  The warm \hi\ component has a Gaussian standard
deviation of 10.1\,$\pm$\,1.2 km s$^{-1}$, while the cold
\hi\ component has a Gaussian standard deviation of 4.2\,$\pm$\,0.9 km
s$^{-1}$.  These values are typical for the warm and cold
\hi\ components seen in many different environments across many
different galaxy types
\citep[e.g.,][]{young96,young97,young03,begum06,deblok06,tamburro09,warren12}.
Assuming a thermalized distribution of particles without turbulence,
these widths correspond to kinetic temperatures of 6,160 K and 1,060 K
for the warm and the cold components, respectively. These temperatures
are similar to what was predicted by \citet{giovanelli13}.

In Figure~\ref{fig: cold hi images} we show the location of the cold
\hi\ detection at 24\arcsec\ resolution in relation to the integrated
\hi, the stellar component, and the single \hii\ region in Leo\,P.
The exact location of the cold \hi\ within the beam is uncertain due
to the unresolved nature of the emission.  However, if molecular
material does exist in Leo P, then the encircled area shown in
Figure~\ref{fig: cold hi images} is the most likely place to observe
it.

A first order estimate of the mass of \hi\ in the cold component is
available by converting the (unresolved) flux density of the cold
component, 3.74 Jy\,Bm$^{-1}$\,\kms, into a mass via the standard
prescription, M$_{\rm HI}$ (\msun) $=$
2.36\,$\times$\,10$^{5}$\,S$_{\rm HI}$\,D(Mpc)$^2$, where S$_{\rm HI}$
is the flux in units of Jy\,\kms.  This yields M$_{\rm
  HI,cold}$ $\ge$ 9000 \msun, which is given as a lower limit,
as more cold \hi\ gas could reside on fine spatial
scales, where our 8\arcsec\ and 4\arcsec\ resolution images lack the
sensitivity to decompose the \hi\ into warm and cold components.  The
ratio of cold to total \hi\ mass in Leo\,P is $\gsim$1\%; this is in
general agreement with, although slightly lower than, the
corresponding mass ratios found for the galaxies studied in
\citet{warren12}.

The detection of a cold neutral ISM component in Leo\,P has important
implications for our understanding of the phase structure of the ISM
in very low-mass halos.  Given that the \hi\ properties of Leo\,P meet
the criteria for classification as a UCHVC
\citep{adams13,giovanelli13}, it represents a critical testbed for
models of the structure of these objects.  Modeling of Leo\,P that is
similar to that performed for Leo\,T in \citet{faerman13} would be
valuable.

\section{Leo P In Context}
\label{section: context}

\subsection{Leo P on the Baryonic Tully-Fisher Relation}
\label{section: btf}

Since its introduction in the seminal \citet{tully77} manuscript, the
Tully-Fisher relation has become a powerful diagnostic of the physical
parameters and distances of galaxies.  As shown in \citet{begum08a},
\citet{stark09}, \citet{trachternach09}, \citet{mcgaugh10a},
\citet{mcgaugh12}, and references therein, the baryonic Tully-Fisher
relationship (BTFR, relating the mass of baryons to the rotation
velocity) appears to hold over many orders of magnitude, from galaxy
cluster scales down to low-mass gas-rich galaxies.

There has been a long standing interest in extending the Tully-Fisher
relationship to the lowest possible rotation velocities.  While the
use of gas-dominated disk galaxies to derive the slope of the BTFR is
well-understood, the extension of the relation to rotation velocities
$<$20 \kms\ is challenging.  As discussed in detail above for Leo\,P,
and in the literature for other extremely low-mass galaxies (e.g.,
Leo\,A, {Young \& Lo 1996}\nocite{young96}), the low rotation
velocities of gas-rich, metal-poor systems make rotation curve work
difficult.  Only a few low-amplitude rotators ($v_{\rm flat}$ $\leq$20
\kms; see discussion in {Begum \etal\ 2008a}\nocite{begum08a}) are
included in the \citet{mcgaugh12} analysis: DDO\,210 ($v_{\rm flat}$
$=$ 17\,$\pm$\,4 \kms), Camelopardalis\,B ($v_{\rm flat}$ $=$
20\,$\pm$\,12 \kms), and UGC\,8215 ($v_{\rm flat}$ $=$ 20\,$\pm$\,6
\kms).  Some previous versions of the BTFR have included dSph galaxies
at the lowest masses; however, \citet{mcgaugh10b} argue that some
dSphs deviate substantially from the BTFR, and suggest that the
environment of these galaxies plays a signficant role in this
departure.

Given these challenges, the addition of even one more low-amplitude
rotator to the BTFR represents a significant step forward.  To this
end, in \citet{giovanelli13} we presented a first estimate of the
location of Leo\,P on the BTFR as calibrated by \citet{mcgaugh12}.
With the present analysis, we are now able to place Leo\,P on this
plot with confidence.  In Figure~\ref{fig: mcgaugh plot} we thus show
two reproductions of the BTFR from \citet{mcgaugh12}. Figure~\ref{fig:
  mcgaugh plot}(a) shows Leo\,P (turquoise square) amongst all of the
galaxies from \citet{mcgaugh12}; red circles are star-dominated
galaxies, green triangles are gas-rich galaxies, and blue squares are
dSph galaxies.  Figure~\ref{fig: mcgaugh plot}(b) shows the same data,
but with the dSph galaxies removed (see discussion above and in
{McGaugh \& Wolf 2010}\nocite{mcgaugh10b} and {Stringer et al.
  2010}\nocite{stringer10}) and the dynamic range compressed for ease
of interpretation.  The shaed grey line shows the $\pm$1$\sigma$
regression to the properties of the gas-dominated galaxies derived in
\citet{mcgaugh12}.  The position of Leo\,P on the BTFR plots is now
well-established, based on the estimate of the rotational velocity
derived in \S~\ref{section: rotation}.  The baryonic mass includes
both atomic gas (corrected for helium) and stars (derived by {McQuinn
  \etal\ 2013a}\nocite{mcquinn13}).  We have made no corrections for
molecular gas mass (see details in {Molter
  \etal\ 2014}\nocite{molter14} and {Warren
  \etal\ 2014}\nocite{warren14}) or for ionized gas mass; note that
\citet{gnedin12} cautions that the ionized gas component is a
substantial mass component in dwarf galaxies.

The position of Leo\,P in the BTFR is now well-understood.  Within the
adopted errorbars on the circular velocity and the baryonic mass,
Leo\,P lies just within the 1$\sigma$ region of the regression derived
by \citet{mcgaugh12} and shown in Figure~\ref{fig: mcgaugh plot}.  Its
baryonic mass is the lowest of any galaxy to date with an estimate of
its circular velocity derived from \hi\ observations.

\subsection{The Association Between Leo\,P and the 14$+$12 Group}
\label{section: 3109}

\citet{mcquinn13} first identified the likely membership of Leo\,P in
the 14$+$12 group, which consists of NGC\,3109, Antlia, Sextans\,A,
Sextans\,B, and GR\,8 \citep{tully02}.  These systems span a range of
distances between 1.25 -- 1.44 Mpc; \citet{mcquinn13} show that
Leo\,P lies at one end of this loose association.  Sextans\,B is the
closest known neighbor to Leo\,P, located 0.47$^{+0.14}_{-0.24}$ Mpc
away in 3D space.  These findings were subsequently verified by
\citet{bellazzini13} and are further discussed in \citet{pawlowski14}.

Given the loose nature of the 14$+$12 association, and the large
distance between Leo\,P and Sextans\,B, it seems unlikely that the
inclusion of Leo\,P in this association has had a measurable impact on
its recent evolution.  In the \hi\ data presented in this work, we do
not find evidence for extended neutral hydrogen gas that would be
indicative of an interaction.  As argued in \citet{mcquinn13}, Leo\,P
remains an ideal example of a metal-poor galaxy that is evolving in
isolation.

\subsection{Leo P Compared To Other Extremely Metal Deficient Galaxies}
\label{section: xmd}

Leo\,P is one of the most extremely metal-deficient galaxies known in
the local universe.  As discussed in \citet{berg12} and
\citet{skillman13}, the number of ``extremely metal deficient''
galaxies (``XMDs'', systems with 12 + log(O/H) $\leq$ 7.65 or Z $\leq$
9\% Z$_{\odot}$ using the Solar oxygen abundance of 12 + log(O/H) =
8.69\,$\pm$\,0.05 from {Asplund \etal\ 2009}\nocite{asplund09})
remains small.  Here we compare the neutral gas properties of Leo\,P
with those of the other systems studied in the \citet{berg12} sample.
The galaxies in that investigation have secure distances, and have
error weighted average nebular oxygen abundances, derived using the
``direct'' method via the auroral [O\,III] $\lambda$4363 \AA\ emission
line, that meet the XMD abundance criterion within observational
errors.  We supplement this sample with characteristics of the
well-studied metal-poor galaxies I\,Zw\,18 and SBS\,0335-052 taken
from the literature.

In Table~\ref{table: xmd params} we summarize relevant physical
parameters of these systems.  Column 1 gives the name(s) of the
galaxy; column 2 gives the error weighted average oxygen abundance on
the 12+log(O/H) scale; column 3 gives the total \hi\ mass of the
galaxy; column 4 gives the stellar mass of the galaxy; column 5 lists
the literature references used to compile the table.  We do not quote
error bars on most values because of the heterogeneity of the datasets
used to accumulate the data; some \hi\ masses are derived via
interferometric measurements while others use single-dish data, etc.
The stellar masses of all systems are derived from Spitzer 4.5 $\mu$m
imaging, except for SBS\,0335-052, I\,Zw\,18, and Leo\,P.  Since the
nebular oxygen abundances of SBS\,0335-052E and SBS\,0335-052W differ,
we treat each component as a separate XMD system.

In Figure~\ref{fig: xmd plot} we plot these parameters as functions
of oxygen abundance for the XMD galaxies.  Figures~\ref{fig: xmd
  plot}(a) and (b) show the \hi\ mass and the stellar mass as
functions of the oxygen abundance.  Leo\,P has the lowest neutral
hydrogen and stellar mass of any of the 22 XMD systems in
Table~\ref{table: xmd params}.  It is important to stress that this is
not simply a proximity effect; some of the systems in this sample are
in the Local Group.  In Figure~\ref{fig: xmd plot}(c) we plot the
M$_{\rm HI}$/M$_{\star}$ ratio as a function of oxygen abundance.  A
Spearman's correlation test reveals a $>$99\% confidence correlation
between higher neutral gas fractions and decreasing oxygen abundances,
although the scatter is large at all metallicities.  It is interesting
to note that even though the \hi\ and stellar masses of Leo\,P are
very low, the M$_{\rm HI}$/M$_{\star}$ ratio is not abnormal compared
to the rest of the sample members.

We are also interested in a comparison of the dynamical properties of
these systems and how these vary as a function of metallicity.
However, such a comparative analysis was not feasible using the data
available in the literature.  Only a few of the XMD systems in
Table~\ref{table: xmd params} have meaningful estimates of the maximum
rotation velocity; further, those that do vary considerably in
technique (e.g., formal rotation curve analysis or representative
rotation via PV slices, differing numbers of beams across the
major-axis, etc.).  Given this inhomogeneity, we do not include this
data in Table~\ref{table: xmd params}, and simply summarize these
properties for a few systems below.

The most complete set of dynamical parameters is available for the few
most metal-poor systems in the sample.  Specifically, for the eight
systems with 12\,$+$\,log(O/H) $<$ 7.5 (again, SBS\,0335-052W and
SBS\,0335-052E are treated as separate systems), dynamical parameters
(including a measurement of rotational velocity and an estimate of the
dynamical mass within the \hi\ radius) are available for five: 
SBS\,0335-052W (V$_{\rm c}$ $\simeq$ 40 \kms;    M$_{\rm dyn}$ $\simeq$ 7.9\,$\times$\,10$^{9}$ \msun; {Ekta \etal\ 2009}\nocite{ekta09}),
Leo\,P         (V$_{\rm c}$ $\simeq$ 15 \kms;    M$_{\rm dyn}$ $>$ 2.6\,$\times$\,10$^{7}$ \msun; this work),
I\,Zw\,18      (V$_{\rm c}$ $\simeq$ 44 \kms;    M$_{\rm dyn}$ $\simeq$ 8.6\,$\times$\,10$^{9}$ \msun; {van Zee \etal\ 1998}\nocite{vanzee98}),
UGC\,5340      (V$_{\rm c}$ $\simeq$ 48-55 \kms; M$_{\rm dyn}$ $\simeq$ 6.0\,$\times$\,10$^{9}$ \msun; {Ekta \etal\ 2008}\nocite{ekta08}),
SBS\,0335-052E (V$_{\rm c}$ $\simeq$ 40 \kms;    M$_{\rm dyn}$ $\simeq$ 7.0\,$\times$\,10$^{9}$ \msun; {Ekta \etal\ 2009}\nocite{ekta09}).
It is interesting to note that the dynamical properties of Leo\,P are
extreme even when compared to those in systems with similar oxygen
abundances: the dynamical mass of Leo\,P (within the \hi\ radius) is
more than two orders of magnitude smaller than that of any other
system.

It is interesting to note that while we do not find evidence for
extended neutral hydrogen gas that would be indicative of infall or of
an interaction in Leo\,P, many (but not all) of the systems in
Table~\ref{table: xmd params} do show such morphological or kinematic
signatures (see references in Table~\ref{table: xmd params}).  The
trend for XMD systems to preferentially harbor disturbed \hi\ disks
was interpreted by \citet{ekta10} as empirical evidence that infall of
pristine gas may result in lower abundances and lower effective
yields.  A holistic dynamical and morphological analysis of all known
XMD galaxies is beyond the scope of this work, but could be a very
promising method with which to establish the fundamental
characteristics (e.g., baryon fraction, mass to light ratios, role of
inall and interactions) of the most metal-poor galaxies.

\section{Discussion and Conclusions}
\label{section: conclusion}

We have presented new {\it Very Large Array} (VLA) and {\it Giant Metrewave
Radio Telescope} (GMRT) \hi\ spectral line observations of the extreme
star-forming galaxy Leo\,P.  This system was discovered in the Arecibo Legacy
Fast ALFA (ALFALFA) survey data products based on its \hi\ characteristics
alone.  Its global neutral gas properties match those of dozens of other
ultra-compact high-velocity clouds (UCHVCs) discovered to date by the ALFALFA
survey; follow-up observations are underway that may reveal more Leo\,P
analogs in the local universe. 

Our combined \hi\ datasets allow us to study the neutral hydrogen in
Leo\,P at high spectral resolution and on spatial scales from tens of
pc to galaxy-wide.  The neutral gas morphology of Leo\,P is smooth on
global scales and the highest \hi\ and stellar surface brightnesses
are cospatial.  A small amount of \hi\ gas extends to the southeast
beyond the extent of the stellar population, while in the northwest
some stars appear to be exterior to the \hi\ of the system at the
present surface brightness sensitivity level.  On progressively finer
spatial scales, the \hi\ gas becomes strongly localized in column
density peaks with \hi\ surface density maxima in excess of 5
\msun\,pc$^{-2}$.  An \hi\ column density maximum of
6.5\,$\times$\,10$^{20}$ cm$^{-2}$ (5.2 \msun\,pc$^{-2}$) is nearly
cospatial (less than one \hi\ beam width, or $\lsim$30 pc) with the
location of the single \hii\ region.

The \hi\ dynamics of Leo\,P are complex.  While ordered rotation is clearly
present from an examination of both the datacubes and of the
intensity-weighted velocity fields at all spatial resolutions, there are
significant irregularities that make modeling the neutral gas disk
challenging.  The velocity dispersion of the disk shows a global value of 8.4
$\pm$ 1.4 \kms, and some regions show dispersions as large as 10-12 \kms.

We derive a low amplitude rotation velocity for Leo\,P: V$_{\rm c}$ $\sim$
15$\pm$5 \kms\ at 60\arcsec\ from the assumed dynamical center.  The implied
dynamical mass is M$_{\rm dyn}$ $>$2.6\,$\times$\,10$^{7}$ \msun\ interior to
the \hi\ radius.  The luminous baryonic mass, calculated as the sum of the
neutral hydrogen mass (corrected for Helium) and the stellar mass, is M$_{\rm
bary}$ $=$ (1.8$^{\rm +0.4}_{\rm -1.2}$)\,$\times$\,10$^6$ \msun.

Using the techniques described in \citet{warren12}, we have shown that the
neutral ISM of Leo\,P contains both a cold and a warm component.  The cold and
warm components have characteristic velocity widths of 4.18\,$\pm$\,0.86 \kms\
and 10.08\,$\pm$\,1.15 \kms, corresponding to kinetic temperatures of
$\sim$1100 K and $\sim$6200 K, respectively.  The cold \hi\ component is
unresolved on 200 pc physical scales.  While the exact location of the cold
\hi\ in Leo\,P is uncertain due to the unresolved nature of the emission, if
molecular material does exist in Leo\,P, then it may be cospatial with this
cold \hi\ gas.

Leo\,P has the lowest neutral hydrogen mass and the lowest stellar mass of any
of the extremely-metal deficient (XMD) galaxies (systems with 12 + log(O/H)
$\leq$ 7.65) studied in \citet{berg12}. Interestingly, its ratio of neutral
hydrogen mass to stellar mass does not stand out amongst the other XMD
galaxies in that work.  The dynamical mass of Leo\,P is more than two orders
of magnitude smaller than that of any other system with comparable
metallicity.  We have placed Leo\,P on the baryonic Tully-Fisher relation as
parameterized in \citet{mcgaugh12}; it represents the most slowly-rotating
gas-rich galaxy studied to date.

\acknowledgements

The authors thank the referee for the helpful review of this manuscript.   
Additionally, we thank the director of the Very Large Array
for the positive review of the discretionary time request for program
12A-456, and for the approval of subsequent observations of Leo\,P in
program 13A-026.  Some of the observations presented in this paper
were obtained using the GMRT which is operated by the National Centre
for Radio Astrophysics (NCRA) of the Tata Institute of Fundamental
Research (TIFR), India.  JMC would like to thank the Instituo
Nazionale di Astrofisica and the Osservatorio Astronomico di Padova
for their hospitality during a productive sabbatical leave.  We thank
R. Koopmann, P. Troischt, and the student members of the Undergraduate
ALFALFA team for conducting the confirming L-band wide observations of
Leo\,P. This investigation has made use of the NASA/IPAC Extragalactic
Database (NED) which is operated by the Jet Propulsion Laboratory,
California Institute of Technology, under contract with the National
Aeronautics and Space Administration, and NASA's Astrophysics Data
System.

JMC is supported by NSF grant 1211683.  The Undergraduate ALFALFA team
is supported by NSF grants AST-0724918, AST-0725267, AST-0725380,
AST-0902211, and AST0903394. The ALFALFA work at Cornell is supported
by NSF grants AST-0607007 and AST- 1107390 to R.G. and M.P.H. and by
grants from the Brinson Foundation. E.A.K.A. was supported by an NSF
predoctoral fellowship during part of this work. K.L.R. is supported
by NSF Faculty Early Career Development (CAREER) award
AST-0847109.  

\clearpage

\clearpage
\bibliographystyle{aj}                                                 


\clearpage
\begin{deluxetable}{lcc}  
\tablecaption{Basic Characteristics of Leo\,P} 
\tablewidth{0pt}  
\tablehead{ 
\colhead{Parameter} &\colhead{Value} &\colhead{Reference}}    
\startdata      
Right ascension (J2000)           & 10$^{\rm h}$ 21$^{\rm m}$ 45.1$^{\rm s}$  &\citet{rhode13} \\
Declination (J2000)               & +18\arcdeg 05\arcmin 17.2\arcsec &\citet{rhode13} \\    
Distance (Mpc)                    & 1.72$^{+0.12}_{-0.40}$ &\citet{mcquinn13} \\
12+log(O/H)                       & 7.17 $\pm$ 0.04 &\citet{skillman13} \\
M$_{\rm V}$ (mag)                   & -9.37$^{+0.15}_{-0.50}$ &\citet{mcquinn13} \\
L$_{\rm H\alpha}$ (erg\,\,s$^{-1}$)     & 6.06$\times$10$^{36}$ &\citet{rhode13} \\
Single-dish S$_{\rm H\,I}$ (Jy \kms)     & 1.31 &\citet{giovanelli13} \\
\hi\ mass M$_{\rm H\,I}$ (\msun)      & 9.5 $\times 10^5$ &\citet{giovanelli13}, this work\\
Stellar mass M$_{\rm \star}$ (\msun)      & 5.7 $\times 10^5$ &\citet{mcquinn13}\\
\enddata     
\label{table: properties}
\begin{small}
\end{small}
\end{deluxetable}   

\clearpage
\begin{footnotesize}
\begin{deluxetable}{lcccc} 
\tablecaption{Properties of Leo\,P and Selected XMD Galaxies}
\tablewidth{0pt}  
\tablehead{ 
\colhead{Galaxy} &\colhead{12$+$log(O/H)} &\colhead{M$_{\rm HI}$} &\colhead{M$_{\star}$}  &\colhead{References\tablenotemark{a}}\\
\colhead{Name(s)} &\colhead{} &\colhead{(10$^6$ M$_{\odot}$)} &\colhead{(10$^6$ M$_{\odot}$)} &\colhead{}}  
\startdata 
SBS 0335-052W           &7.12\,$\pm$\,0.03   &580        &12.0                &1\nocite{izotov05}, 2\nocite{ekta09}     \\
Leo\,P                  &7.17\,$\pm$\,0.04   &0.95       &0.57                &3, 4\nocite{skillman13}, 5\nocite{mcquinn13}\\ 
I\,Zw\,18               &7.17\,$\pm$\,0.04   &230        &90                  &6\nocite{skillman93}, 7\nocite{izotov99}, 8\nocite{vanzee98}, 9\nocite{aloisi07}, 10\nocite{fisher14}\\ 
UGC\,5340 (DDO\,68)     &7.20\,$\pm$\,0.05   &1,000      &93.3                &11\nocite{berg12},12\nocite{ekta08}     \\         
UGCA\,292 (CVn\,I\,dwA) &7.30\,$\pm$\,0.03   &40         &4.79                &11\nocite{berg12},13\nocite{ott12}      \\          
Leo\,A (DDO\,69)        &7.30\,$\pm$\,0.05   &6.9        &3.80                &11\nocite{berg12}, 14\nocite{hunter12}     \\          
SBS 0335-052E           &7.33\,$\pm$\,0.01   &420        &41.0                &15\nocite{izotov97},2\nocite{ekta09}, 16\nocite{pustilnik04}    \\  
CGCG\,269-049           &7.47\,$\pm$\,0.05   &12         &0.794               &11\nocite{berg12}, 13\nocite{ott12}      \\ 
Sextans\,B (DDO\,70)    &7.53\,$\pm$\,0.05   &42         &30.9                &11\nocite{berg12}, 13\nocite{ott12}      \\          
UGC\,6817 (DDO\,99)     &7.53\,$\pm$\,0.05   &47         &9.33                &11\nocite{berg12}, 13\nocite{ott12}      \\ 
Sextans\,A (DDO\,75)    &7.54\,$\pm$\,0.06   &62         &12.0                &11\nocite{berg12}, 13\nocite{ott12}      \\          
UGC\,4483               &7.56\,$\pm$\,0.03   &33         &2.63                &11\nocite{berg12}, 13\nocite{ott12}      \\ 
NGC\,4163               &7.56\,$\pm$\,0.14   &9.3        &40.7                &11\nocite{berg12}, 13\nocite{ott12}      \\ 
UGC\,668 (IC\,1613)     &7.62\,$\pm$\,0.05   &34         &13.8                &11\nocite{berg12}, 14\nocite{hunter12}      \\ 
UGC\,8091 (GR\,8)       &7.65\,$\pm$\,0.06   &5.9        &3.55                &11\nocite{berg12}, 13\nocite{ott12}       \\ 
UGC\,7605               &7.66\,$\pm$\,0.11   &22         &13.2                &11\nocite{berg12}, 17\nocite{begum08b}      \\ 
UGC\,521                &7.67\,$\pm$\,0.05   &290        &92.1                &11\nocite{berg12}, 18\nocite{springob05}      \\ 
NGC\,3741               &7.68\,$\pm$\,0.05   &81         &11.2                &11\nocite{berg12}, 13\nocite{ott12}      \\ 
UGC\,4278  (IC\,2233)   &7.69\,$\pm$\,0.05   &160        &316                 &11\nocite{berg12}, 19\nocite{stark09}      \\             
UGC\,695                &7.69\,$\pm$\,0.12   &72         &120                 &11\nocite{berg12}, 20\nocite{nichols11}      \\            
UGC\,5923 (Mrk\,1264)   &7.79\,$\pm$\,0.14   &42         &195                 &11\nocite{berg12}, 21\nocite{haynes11}      \\      
\enddata
\label{table: xmd params}
\begin{footnotesize}
\tablenotetext{a}{References: 
1 - \citet{izotov05}; 
2 - \citet{ekta09};
3 - This work; 
4 - \citet{skillman13}; 
5 - \citet{mcquinn13};
6 - \citet{skillman93}; 
7 - \citet{izotov99}; 
8 - \citet{vanzee98}; 
9 - \citet{aloisi07}; 
10 - \citet{fisher14};
11 - \citet{berg12};
12 - \citet{ekta08};    
13 - \citet{ott12};    
14 -  \citet{hunter12};     
15 - \citet{izotov97};
16 - \citet{pustilnik04};    
17 - \citet{begum08b};    
18 - \citet{springob05};    
19 - \citet{stark09};            
20 - \citet{nichols11};              
21 - \citet{haynes11}}
\end{footnotesize}
\end{deluxetable}
\end{footnotesize}

\clearpage
\begin{figure}
\epsscale{0.9} 
\plotone{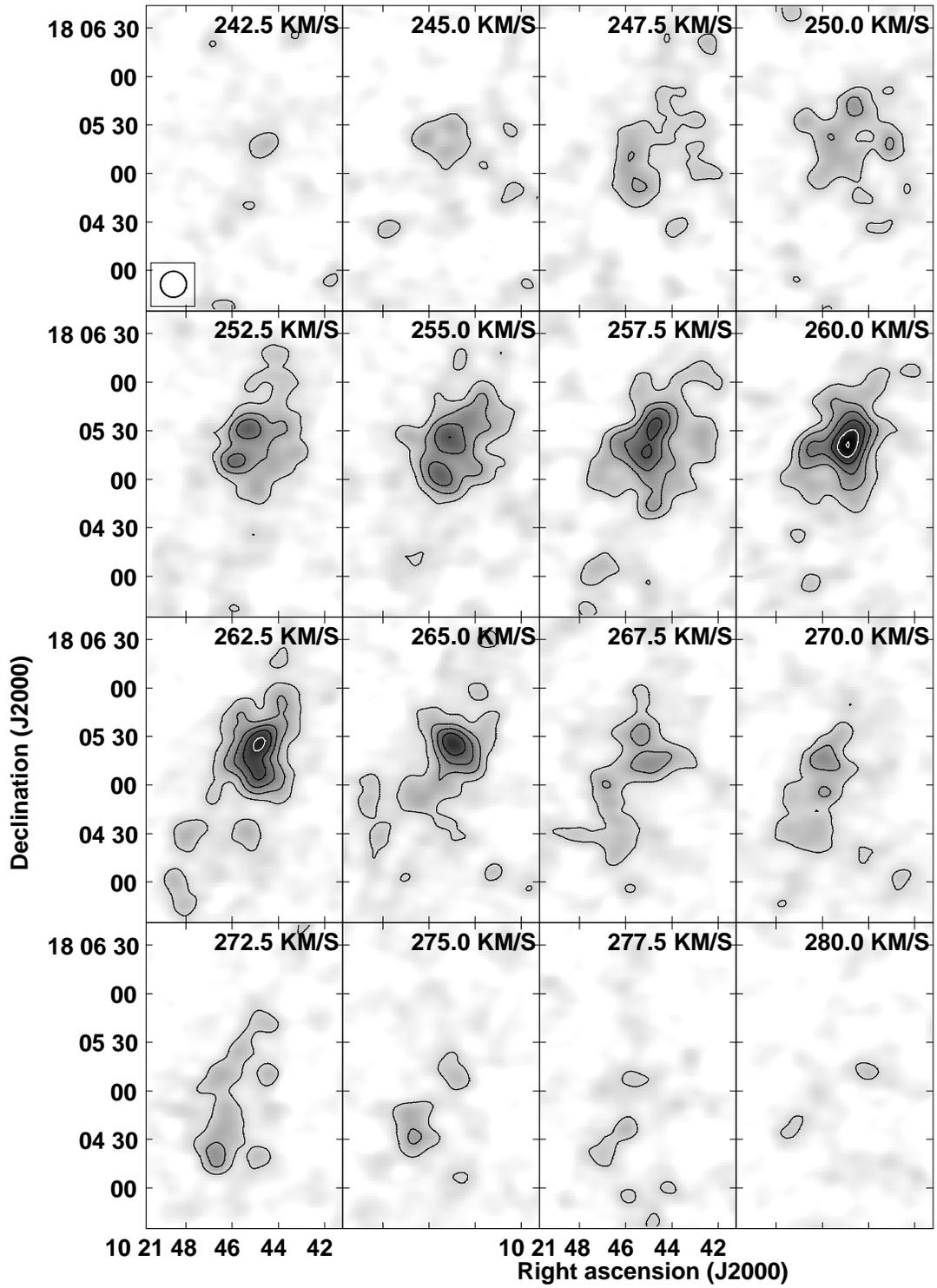}
\caption{Channel maps of \hi\ emission in Leo\,P at
  16\arcsec\ resolution (beam shown in upper left panel).  The radial
  velocity is shown in each panel, and the contours show \hi\ at flux
  levels of (1, 2, 3, 4, 5, 6) mJy\,Bm$^{-1}$; the noise level in the
  cube is 0.51 mJy\,Bm$^{-1}$.}
  \label{fig: chmaps} 
\end{figure}  

\clearpage
\begin{figure}
\epsscale{1.0} 
\plotone{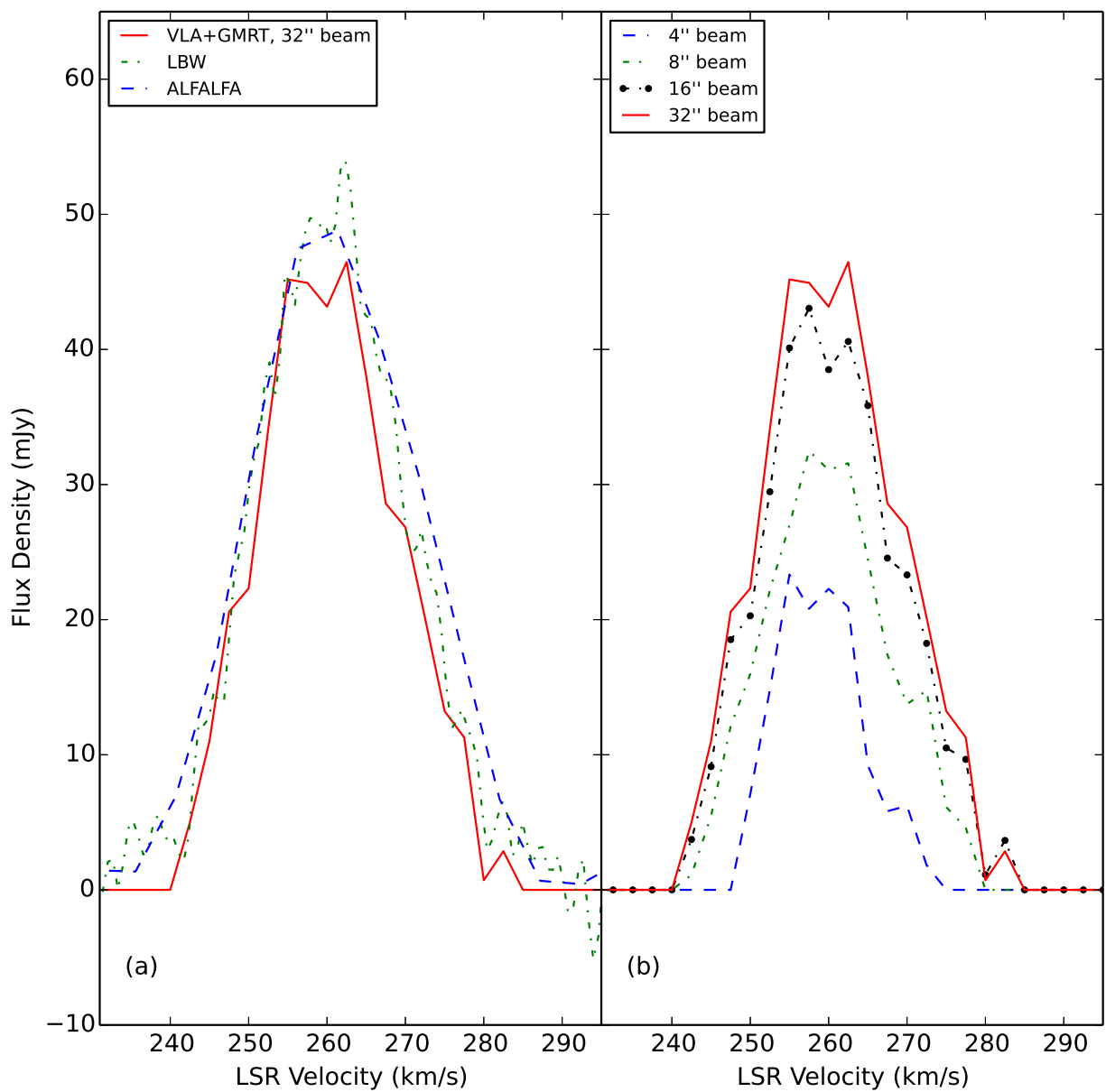}
\caption{ (a): Integrated spectra of Leo\,P from the VLA+GMRT
  observations at 32\arcsec\ spatial resolution (solid red line), from
  the ALFALFA data products (blue dashed line), and from the follow-up
  Arecibo observation with the LBW receiver at Arecibo (green dashed
  line). (b) Integrated spectra of Leo\,P from VLA+GMRT observations
  at 4\arcsec, 8\arcsec, 16\arcsec, and 32\arcsec\ spatial resolutions
  with 2.5 \kms\ spectral resolution. As expected, due to masking, the
  flux density of the source decreases with increasing spatial
  resolution.}
\label{fig: spectra} 
\end{figure}  

\clearpage
\begin{figure}
\epsscale{1.0} 
\plotone{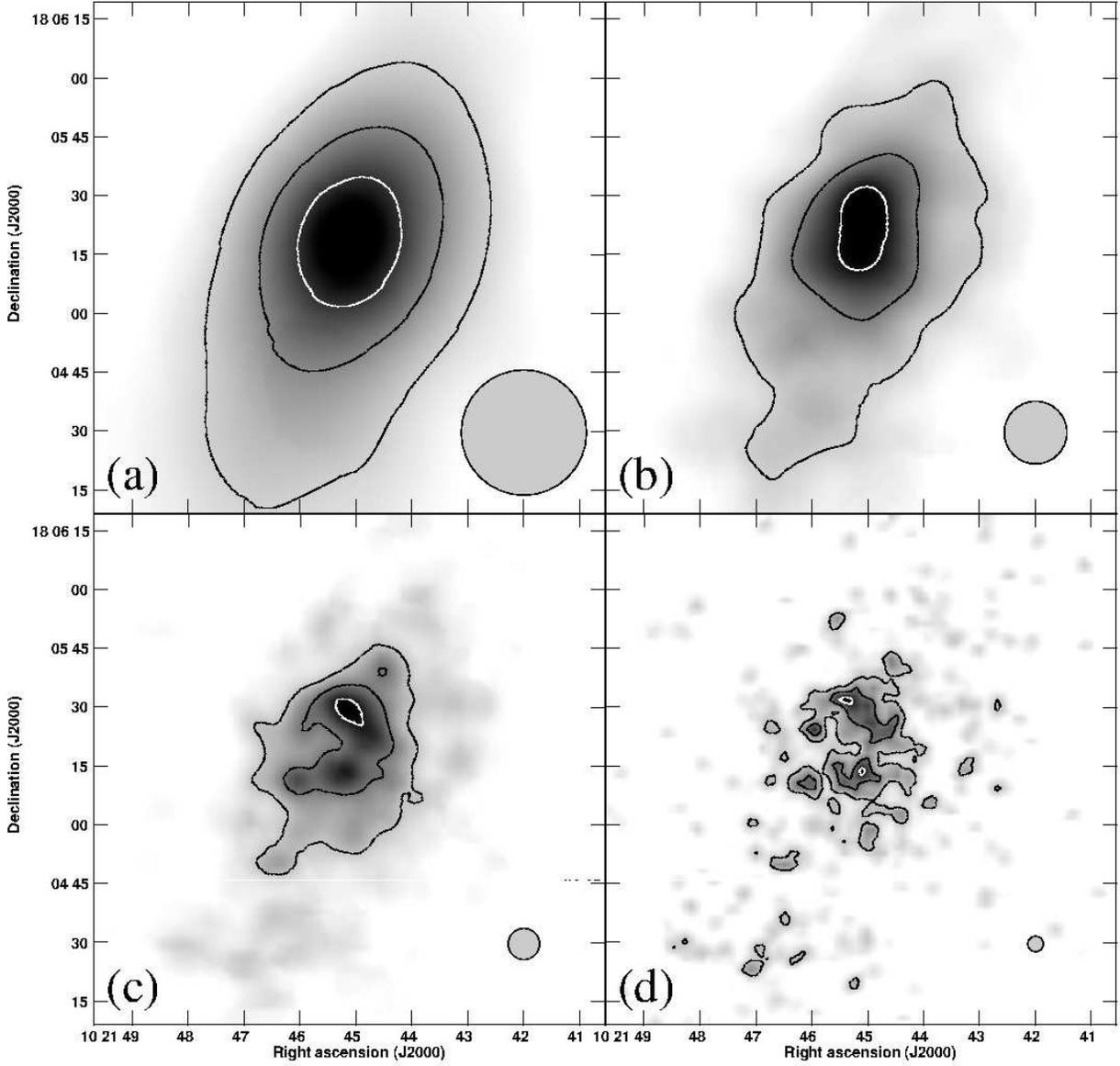}
\caption{\hi\ column density images of Leo\,P at 32\arcsec\ (a),
  16\arcsec\ (b), 8\arcsec\ (c), and 4\arcsec\ (d) resolutions.  The
  images are shown in greyscale; the column density contours, in units
  of 10$^{20}$ cm$^{-2}$, are at levels of (0.75, 1.5, 2.5), (1.0,
  2.25, 3.5), (1.5, 3, 4.5), and (2, 4, 6) in panels (a), (b), (c),
  and (d), respectively.  The beam sizes are shown by filled circles
  in the lower right of each panel. These contours are overlaid on
  optical images of Leo\,P in Figure~\ref{fig: lbt}.}
\label{fig: coldens} 
\end{figure}  

\clearpage
\begin{figure}
\epsscale{1.0} 
\plotone{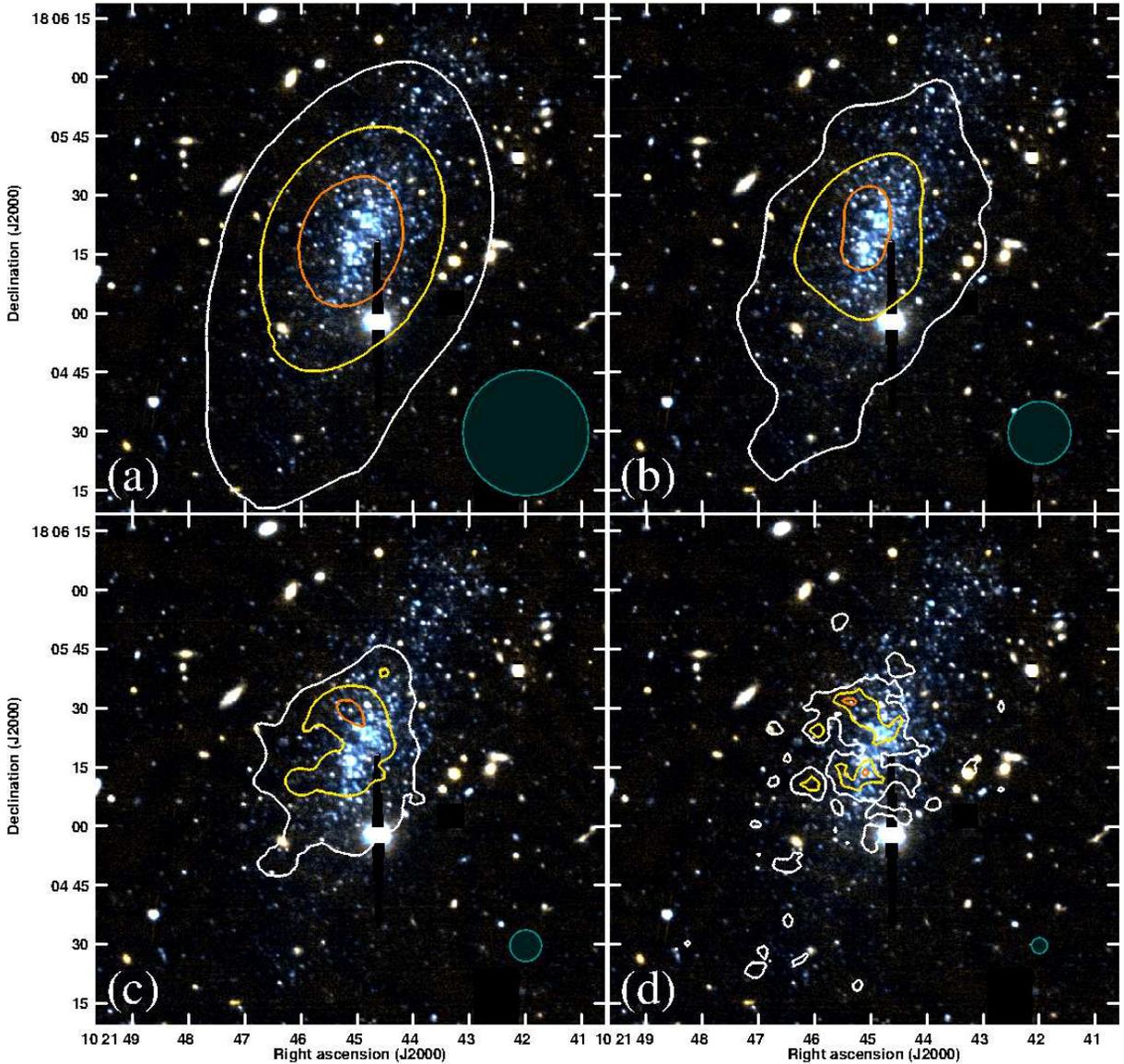}
\caption{\hi\ column density images of Leo\,P at 32\arcsec\ (a),
  16\arcsec\ (b), 8\arcsec\ (c), and 4\arcsec\ (d) resolutions,
  overlaid as contours on the optical 3-color image derived from the
  LBT images presented in \citet{mcquinn13}.  \hi\ column density
  contours are color-coded from lowest (white) to highest (orange) in
  each panel.  The column density contours, in units of 10$^{20}$
  cm$^{-2}$, are at levels of (0.75, 1.5, 2.5), (1.0, 2.25, 3.5),
  (1.5, 3, 4.5), and (2, 4, 6) in panels (a), (b), (c), and (d),
  respectively.  The beam sizes are shown by filled circles in the
  lower right of each panel.  Note that the \hi\ column density
  exceeds 6\,$\times$\,10$^{20}$ cm$^{-2}$ in close proximity to the
  single \hii\ region (see Figure~\ref{fig: lbt zoom} for details).
  Bleed trails due to bright Milky Way foreground stars have been
  removed manually.}
\label{fig: lbt} 
\end{figure}  

\clearpage
\begin{figure}
\epsscale{1.0} 
\plotone{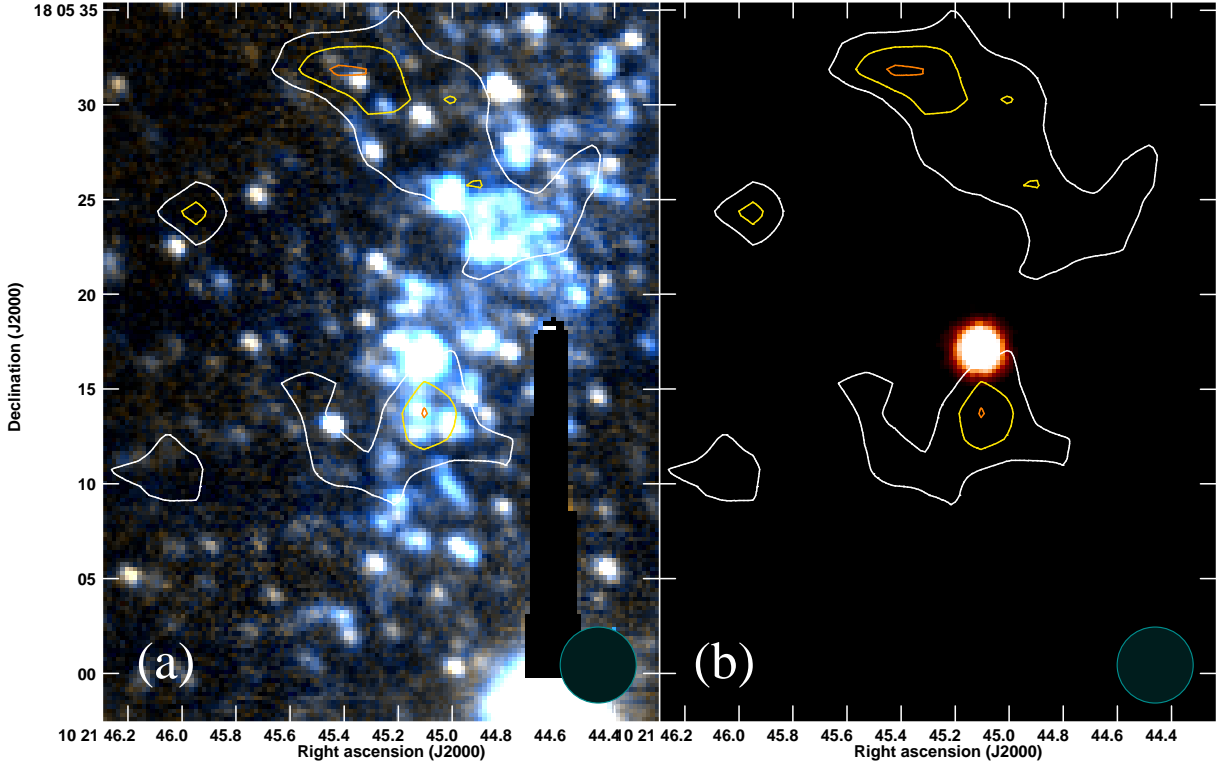}
\caption{\hi\ column density contours at 4\arcsec\ resolution,
  overlaid on LBT 3-color (a) and on WIYN 2.1\,m continuum subtracted
  H$\alpha$ (b) images of Leo\,P; the field of view is smaller than
  that shown in Figure~\ref{fig: lbt}.  Contours are shown at levels
  of (4.0, 5.25, 6.5)\,$\times$\,10$^{20}$ cm$^{-2}$.  The single
  \hii\ region is in very close proximity to an \hi\ column density
  maximum (6.5\,$\times$\,10$^{20}$ cm$^{-2}$).  A second column
  density peak, which is not currently associated with ongoing star
  formation, is located slightly offset from the main stellar
  component.  The beam size is shown by a filled circle in the lower
  right of each panel.  Bleed trails due to bright Milky Way
  foreground stars have been removed manually.}
\label{fig: lbt zoom} 
\end{figure}  

\clearpage
\begin{figure}
    \epsscale{0.55} 
    \plotone{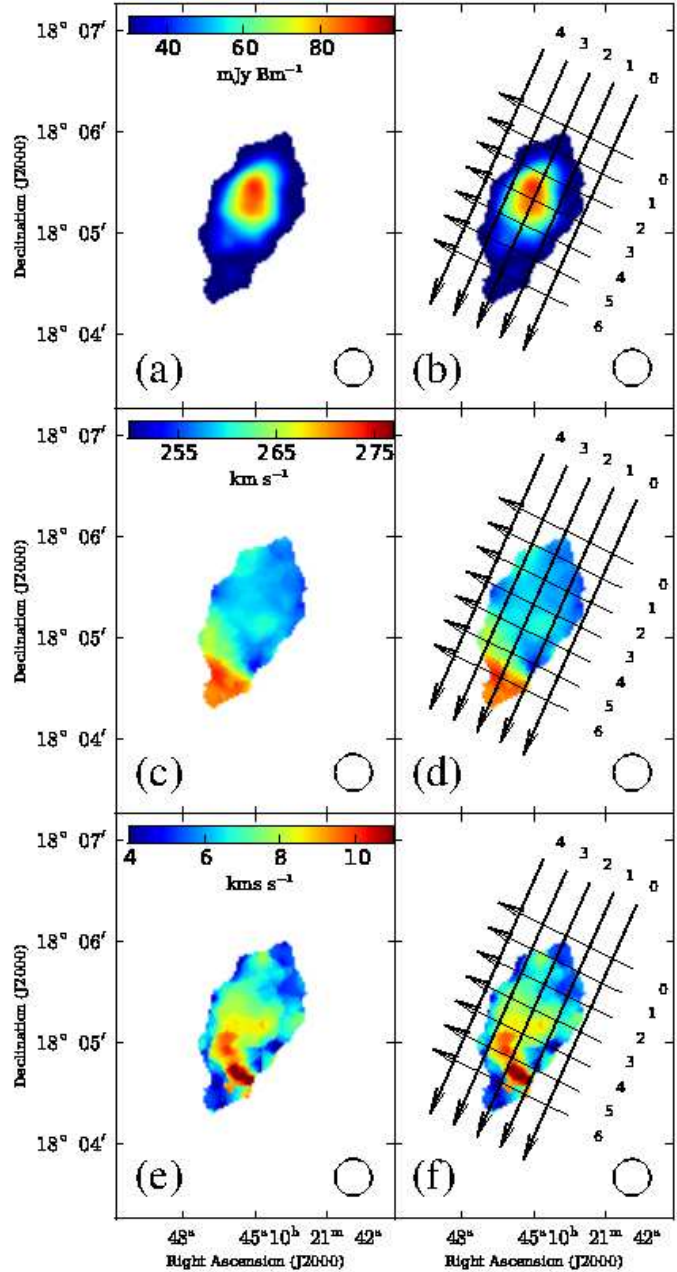}
    \caption{(a) + (b): Map of integrated \hi\ emission from the
      16\arcsec\ resolution cube. The beamsize is represented by the
      circle in the lower-right of each image. (c) + (d):
      Intensity-weighted \hi\ velocity fields from the same cube. (e)
      + (f): Intensity-weighted \hi\ velocity dispersion field from
      the same cube. Panels (b), (d), and (f) are overlaid with arrows
      representing the direction and length of the position-velocity
      slices taken. The arrows leading from the northwest to the
      southeast represent the major-axis position-velocity slices
      which sample 2.7\arcmin.  The arrows leading from the southwest
      to the northeast represent the minor-axis position-velocity
      slices which sample 1.5\arcmin. Major-axis slice 2 and
      minor-axis slice 3 are centered on the peak column density.
      Each slice along the major-axis and minor-axis is separated by a
      beam width; thus each slice is independent of the adjacent
      slice. The numbers at the origins of the arrows correspond to
      the frames in Figure~\ref{fig: pvslices}. }
    \label{fig: pvslice maps} 
\end{figure}

\clearpage
\begin{figure}
\epsscale{1.0}
\plotone{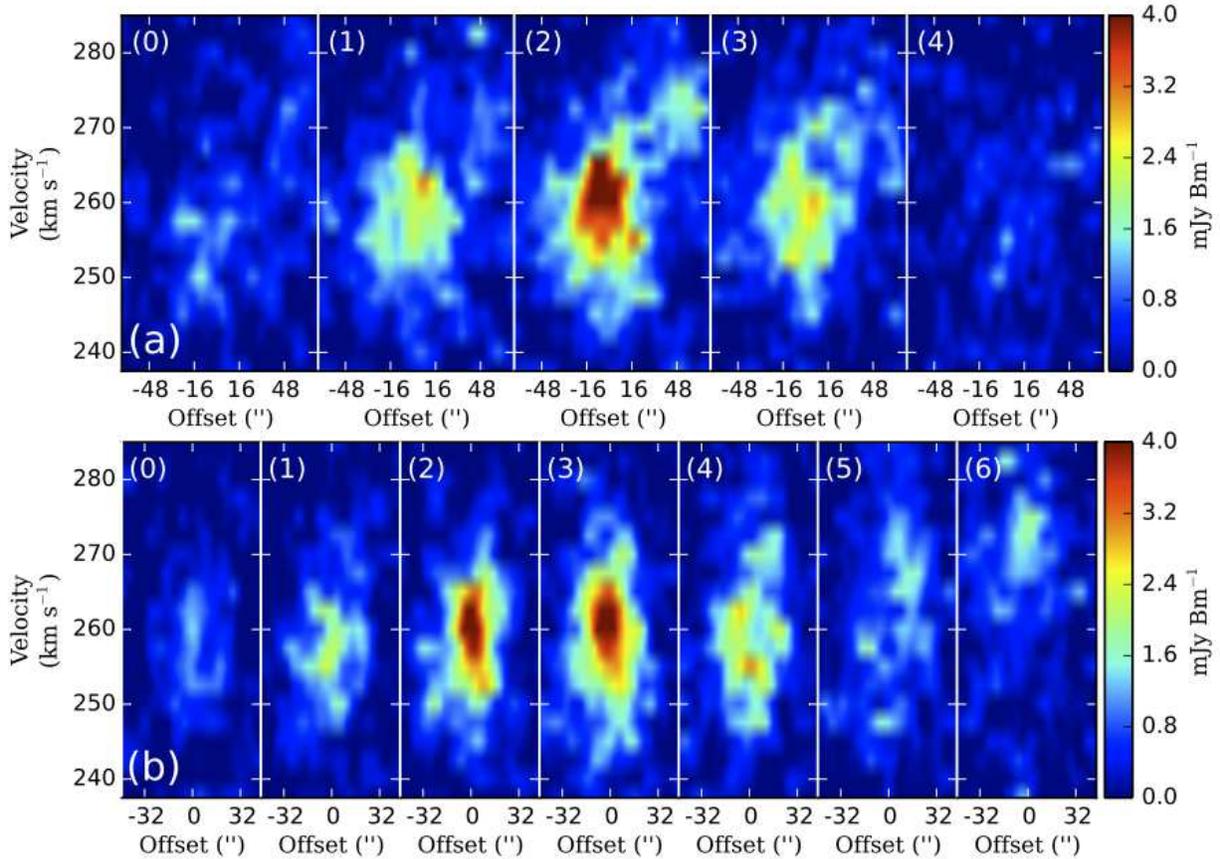}
\caption{(a) Major-axis position-velocity slices of the neutral gas in
  Leo\,P separated by a beam width at 16\arcsec\ resolution. (b) Same
  as (a) except for the minor-axis. Each slice averages pixels over a
  beamsize, and is completely independent from an adjacent slice. See
  Figure~\ref{fig: pvslice maps} for an illustration of the slices
  taken. The apparent slope in the velocity distribution between the
  minor-axis slices can be interpreted as rotation. The velocity
  dispersion as shown by the extended emission across many velocities
  is only slightly lower than any apparent rotational velocity seen in
  Leo\,P.}
  \label{fig: pvslices} 
\end{figure}  

\clearpage
\begin{figure}[H]
    \epsscale{1}
    \plotone{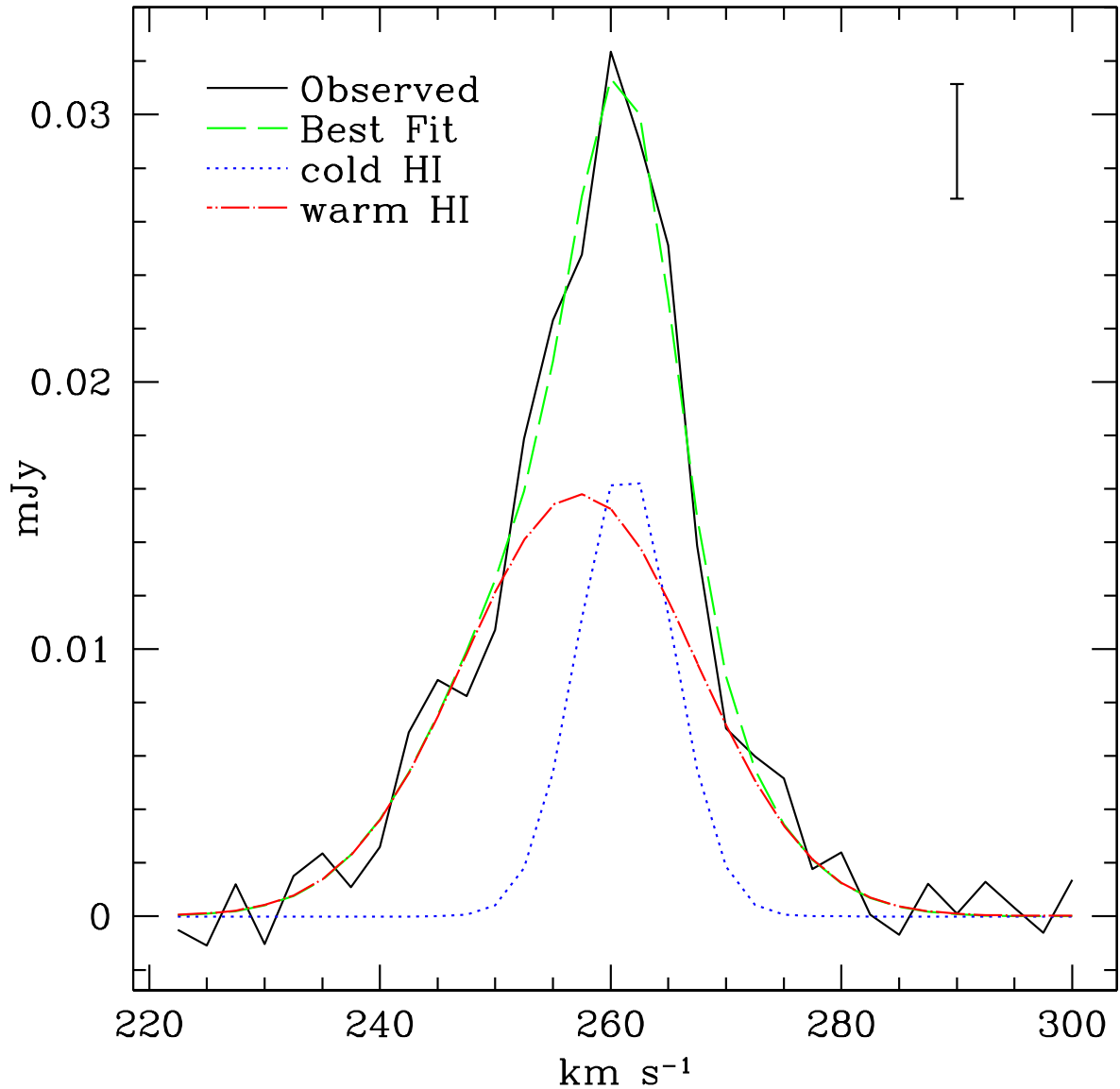}
    \caption{The spectrum at the location of the peak cold
      \hi\ component detected in Leo\,P at 24\arcsec\ (200 pc)
      resolution. The cold \hi\ component represents 30\% of the total
      \hi\ flux in the peak location.  The error bar in the upper
      right represents the error of the spectrum derived from the
      non-flux-rescaled cube of 0.64 \mjpbeam\ (0.002 mJy).  The warm
      \hi\ component has a velocity width of 10.08 $\pm$ 1.15
      \kms\ and the cold \hi\ component has a velocity width of 4.18
      $\pm$ 0.86 \kms. Assuming a thermalized distribution of
      particles without turbulence, these widths correspond to kinetic
      temperatures of 6160 K and 1060 K for the warm and the cold
      components, respectively.}
    \label{fig: cold hi spectrum} 
\end{figure}

\clearpage
\begin{figure}[H]
    \epsscale{1}
    \plotone{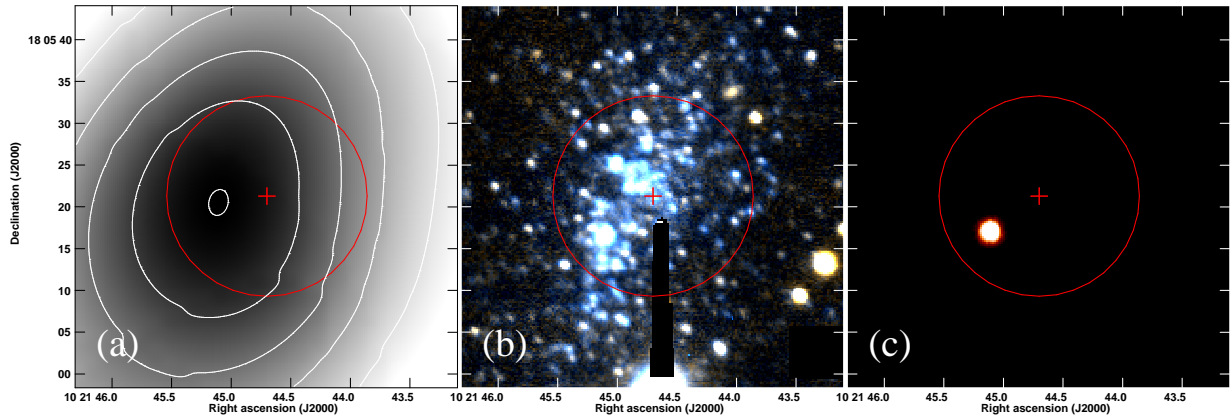}
    \caption{\hi\ image at 24\arcsec\ resolution (a), LBT color image
      (b), and H$\alpha$ image (c), each overlaid with a 200 pc
      diameter red circle representing the location of the
      (unresolved) cold \hi\ component.  The contours on the
      \hi\ image range from (0.8--3.8)\,$\times$\,10$^{20}$ cm$^{-2}$
      in steps of 5\,$\times$\,10$^{19}$ cm$^{-2}$. The exact location
      of the cold \hi\ within the 24\arcsec\ beam is uncertain due to
      the unresolved nature of the emission.}
    \label{fig: cold hi images} 
\end{figure}

\clearpage
\begin{figure}[H]
    \epsscale{0.95}
    \plotone{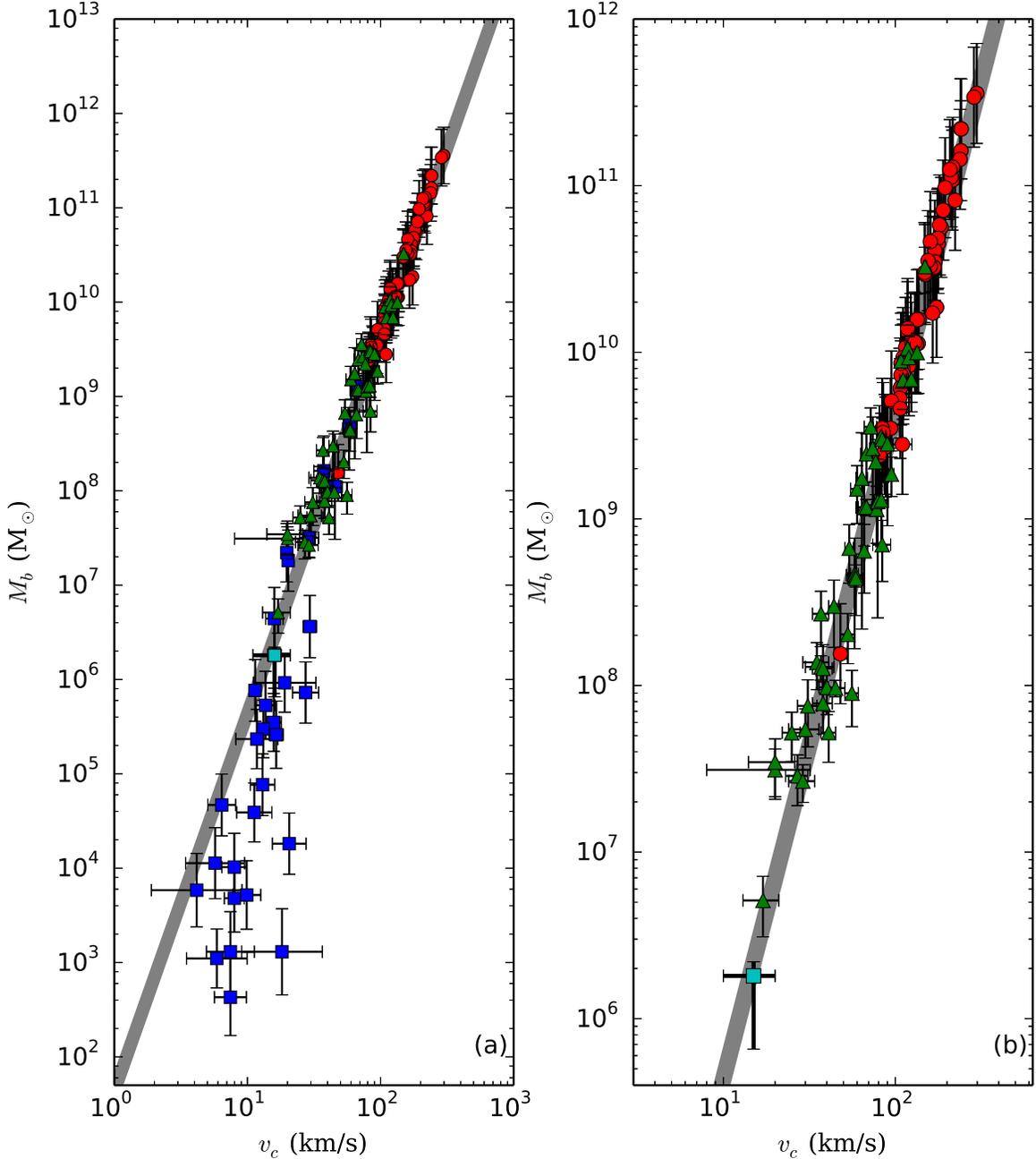}\vspace{-0.5 cm}
    \caption{Reproductions of the baryonic Tully-Fisher relation
      (BTFR) from \citet{mcgaugh12}. Baryonic masses are plotted
      against the circular velocities of galaxies; red circles are
      star-dominated galaxies, green triangles are gas-rich galaxies,
      and blue squares are dSph galaxies.  The shaded grey line shows
      the $\pm$1$\sigma$ regression to the properties of the
      gas-dominated derived in \citet{mcgaugh12}. The turquoise square
      represents Leo\,P.  Panel (a) shows all galaxies from
      \citet{mcgaugh12}, while panel (b) removes the dSph systems and
      compresses the dynamic range for ease of interpretation.  The
      data point for Leo\,P is plotted at the rotational velocity of
      15 \kms\ derived in \S~\ref{section: rotation}. Leo\,P is the
      slowest rotating gas-rich galaxy studied to date.  The position
      of Leo\,P in these plots is consistent with the preliminary
      estimate presented in \citet{giovanelli13}.}
    \label{fig: mcgaugh plot} 
\end{figure}

\clearpage
\begin{figure}[H]
    \epsscale{1.0}
    \plotone{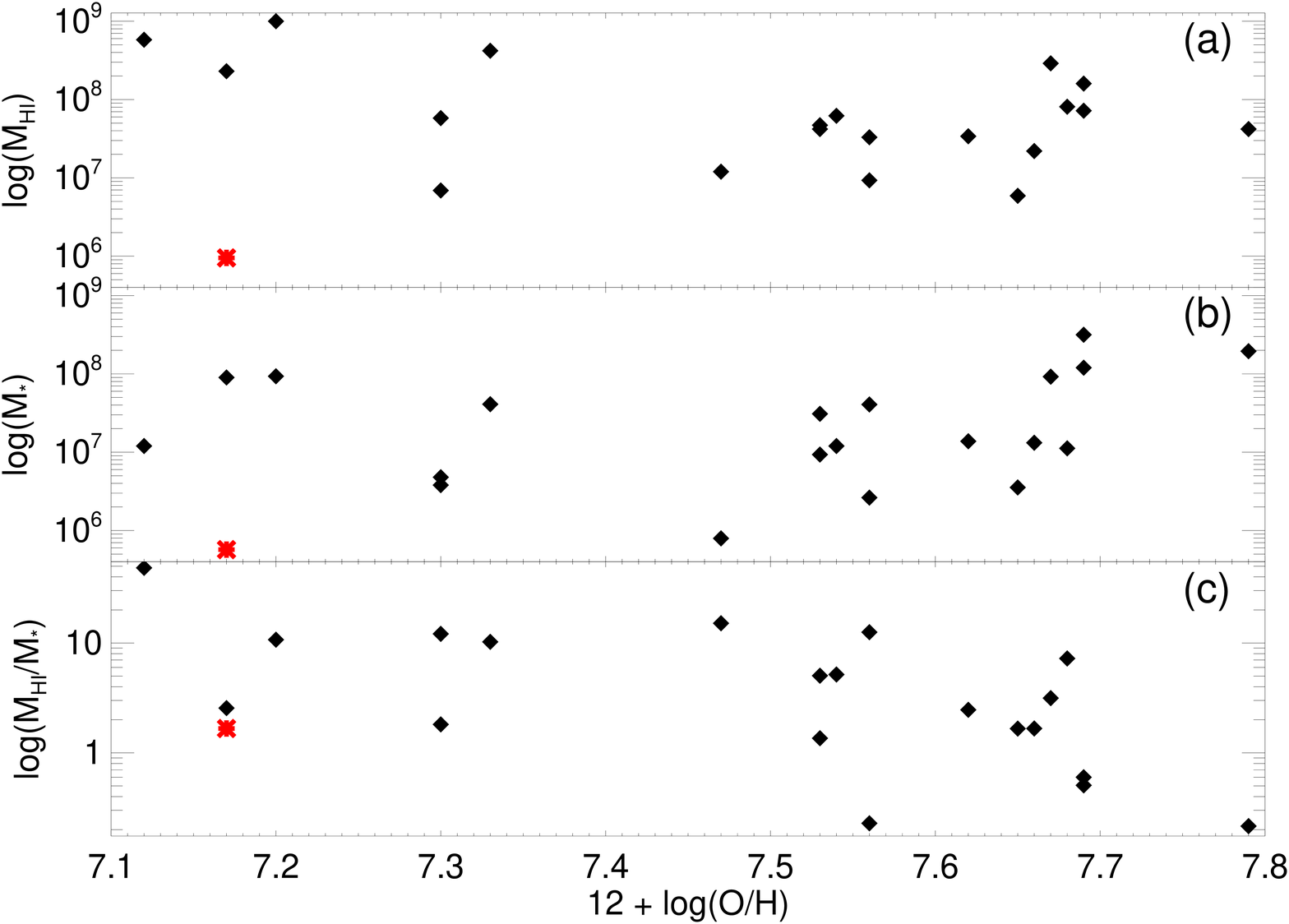}
    \caption{The \hi\ mass (a), stellar mass (b), and the ratio of \hi\ mass to
    stellar mass (c), plotted versus oxygen abundance for selected XMD
galaxies.  Leo\,P is shown in red in each panel.  The parameters of each galaxy
are found in Table~\ref{table: xmd params}.}
      \label{fig: xmd plot} 
\end{figure}

\end{document}